\documentclass[journal = aamick]{achemso}

\usepackage{filecontents}
\usepackage{amssymb}

\usepackage{graphicx}% Include figure files
\usepackage{dcolumn}% Align table columns on decimal point
\usepackage{bm}% bold math
\usepackage{upgreek}

\usepackage{makecell}
\usepackage{array}
\newcolumntype{?}{!{\vrule width 0.8pt}}

\usepackage{color, colortbl}
\definecolor{Gray}{gray}{0.85}

\usepackage{xr}
\author{Min-Hsuan Peng}
\affiliation{Department of Physics, National Taiwan Normal University, Taipei 116, Taiwan}
\altaffiliation{Contributed equally to this work}
\author{Ching-Yang Pan}
\affiliation{Department of Physics, National Taiwan Normal University, Taipei 116, Taiwan}
\altaffiliation{Contributed equally to this work}
\author{Hao-Xuan Zheng}
\affiliation{Department of Physics, National Sun Yat-Sen University, Kaohsiung 804, Taiwan}
\author{Ting-Chang Chang}
\affiliation{Department of Physics, National Sun Yat-Sen University, Kaohsiung 804, Taiwan}
\author{Pei-hsun Jiang}
\affiliation{Department of Physics, National Taiwan Normal University, Taipei 116, Taiwan}
\email{pjiang@ntnu.edu.tw}

\title{Dynamic Behaviors and Training Effects in TiN/Ti/HfO$_x$/TiN Nanolayered Memristors with Controllable Quantized Conductance States: Implications for Quantum and Neuromorphic Computing Devices}

\begin{document}

\begin{abstract}
Controllable quantized conductance states of TiN/Ti/HfO$_x$/TiN memristors are realized with great precision through a pulse-mode reset procedure, assisted with analytical differentiation of the condition of the set procedure, which involves critical monitoring of the measured bias voltage. An intriguing training effect that leads to faster switching of the states is also observed during the operation. Detailed analyses on the low- and high-resistance states under different compliance currents reveal a complete picture of the structural evolution and dynamic behaviors of the conductive filament in the HfO$_x$ layer. This study provides a closer inspection on the quantum-level manipulation of nanoscale atomic configurations in the memristors, which helps to develop essential knowledge about the design and fabrication of the future memristor-based quantum devices and neuromorphic computing devices.
\\
\\
Keywords: \textit{HfO$_2$, filament, resistive random-access memory (RRAM), memristor, oxygen vacancy, resistive switching, conductance quantization, training effect}
\end{abstract}

\maketitle

\section{Introduction}\label{intro}

Memristors with high scalability, low power consumption, and multilevel switching is one of the promising candidates for artificial synapses in neuromorphic computing to replace the conventional von-Neumann architecture \cite{Chua1971,Yang2013,Milo2020}.  Linear conductance of a memristor in a wide voltage range is pursued for the purpose of implementing vector-matrix multiplication in conductance programming for memristor arrays \cite{Li2015a,Merced-Grafals2016,Sun2018}. Nonlinear memristor dynamics due to intrinsic conduction mechanisms \cite{Pickett2009,Menzel2011,Strachan2013} are therefore one of the key challenges to build memristor-based dot-product engines.

In the mean time, quantized conduction in memristor-based devices is being introduced to this field for its great potential for high-density data storage through multilevel switching, and for analog synaptic weight update in effective training of the artificial neural networks \cite{Lim2018,Chen2020}. While implementations of quantum neuromorphic computing platforms with quantum memristors are being proposed \cite{Markovic2020,Pfeiffer2016}, realization of these architectures remains difficult because conductance quantization of the memristors suffers significant instability in terms of endurance and tuning accuracy, which includes large half-widths in the histogram of the quantized conductance \cite{Li2015,Xue2019}. The occurrence of conductance quantization seems unstable and random even when an optimal condition is used in the measurement \cite{Ohno2011,Yi2016,Xie2020}. The mechanism that guarantees conductance quantization in a memristor remains a mystery. The detailed atomic dynamics of the conductive filaments in the memristors is not yet fully explored and therefore demands more research.

In this letter, we have made in-depth investigations on conductance characteristics of bipolar TiN/Ti/HfO$_x$/TiN  valence-change memristors (VCMs), aiming to look into the instability issue of the quantized conductance. The dynamics of the set procedure is observed to be decisive for the quantization performance in the reset procedure. Detailed analyses on the low-resistance state (LRS) have also been conducted to explicitly explore the electrical characteristics associated with the nanoscale atomic structure of the conductive filament in the HfO$_x$ layer. With better understanding of the atomic dynamic behaviors of the conductive filament, we are able to perform a precise control of the quantized conductance states of the memristors.

\section{Device Fabrication and Measurement Methods}\label{sec:device}

\begin{figure}[h]
	\centering
\includegraphics[width=0.5\textwidth]{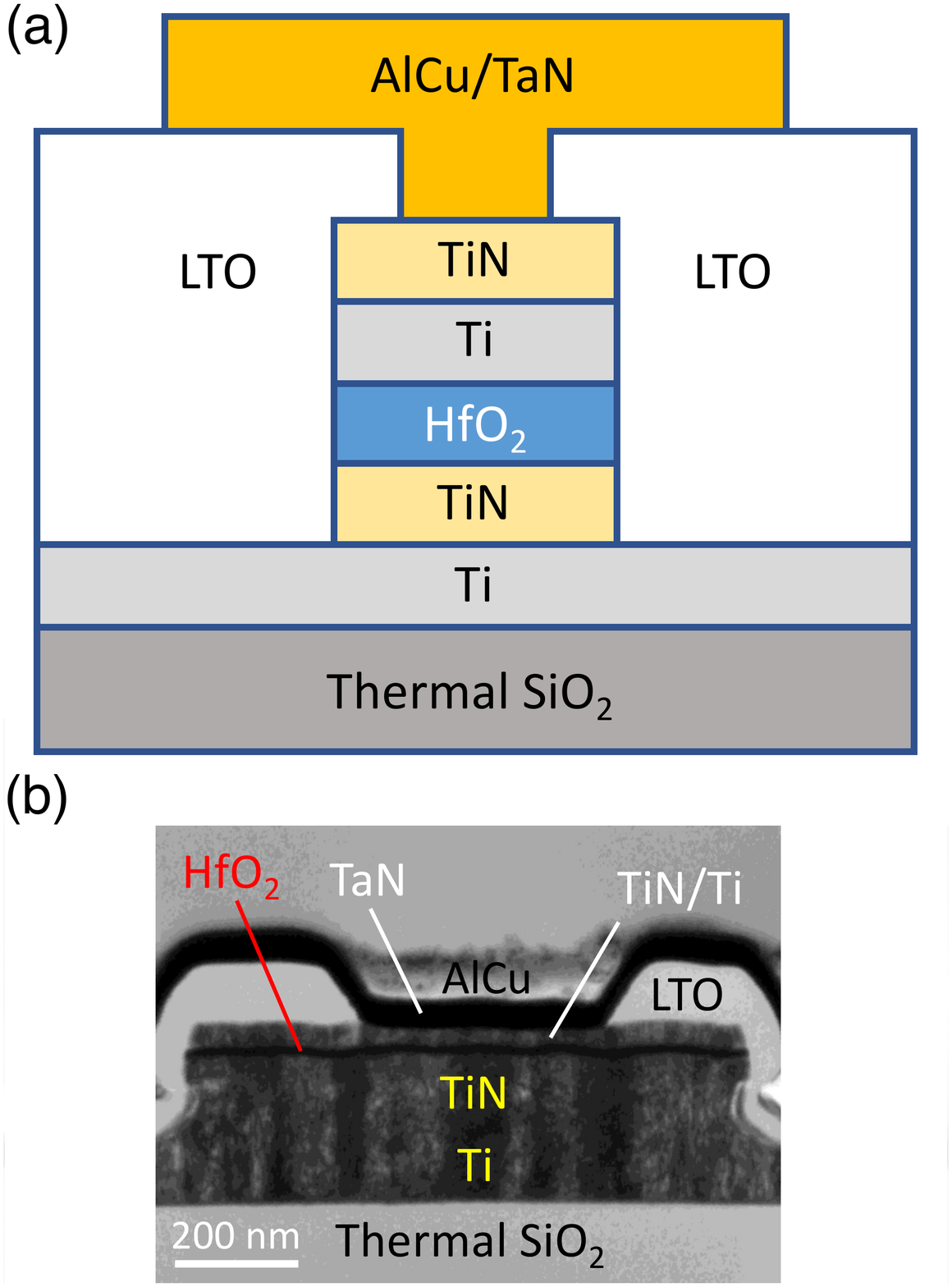}
	\caption{(a) Schematic drawing (not to scale) and (b) the TEM image of the cross-sectional view of the TiN/Ti/HfO$_x$/TiN memristor.}
	\label{device}
\end{figure}

Fig.~\ref{device} shows the device layout of the TiN/Ti/HfO$_x$/TiN memristor. A 300-nm SiO$_2$ layer was grown via wet oxidation on a lightly doped p-type Si(100) substrate. The SiO$_2$ layer serves as an insulating layer between the Si substrate and the bottom electrodes, which were formed by depositing TiN (50 nm)/Ti (50 nm) using radio-frequency (rf) sputtering. After that, a 10-nm switching layer of HfO$_2$ was formed with atomic layer deposition, followed by a deposition of TiN (40 nm)/Ti (10 nm) layer as top electrodes. Lithography and inductively-coupled-plasma (ICP) etching were then used to define the active cell area. Then, a low-temperature oxide (LTO) SiO$_2$ layer was deposited, followed by lithography and ICP etching to create via-hole structures in LTO. Finally, AlCu/TaN contacts for the electrodes were formed via lithography and rf sputtering. The electrical characteristics are measured at room temperature using Keithley 2400 and Agilent B1500A. The bias voltages are applied to the top electrodes as the bottom electrodes are grounded during electrical measurements.

\section{Results and Discussion}
\subsection{Measurements with the DC Voltage Sweep Mode}

\begin{figure}[h]
	\centering
\includegraphics[width=0.7\textwidth]{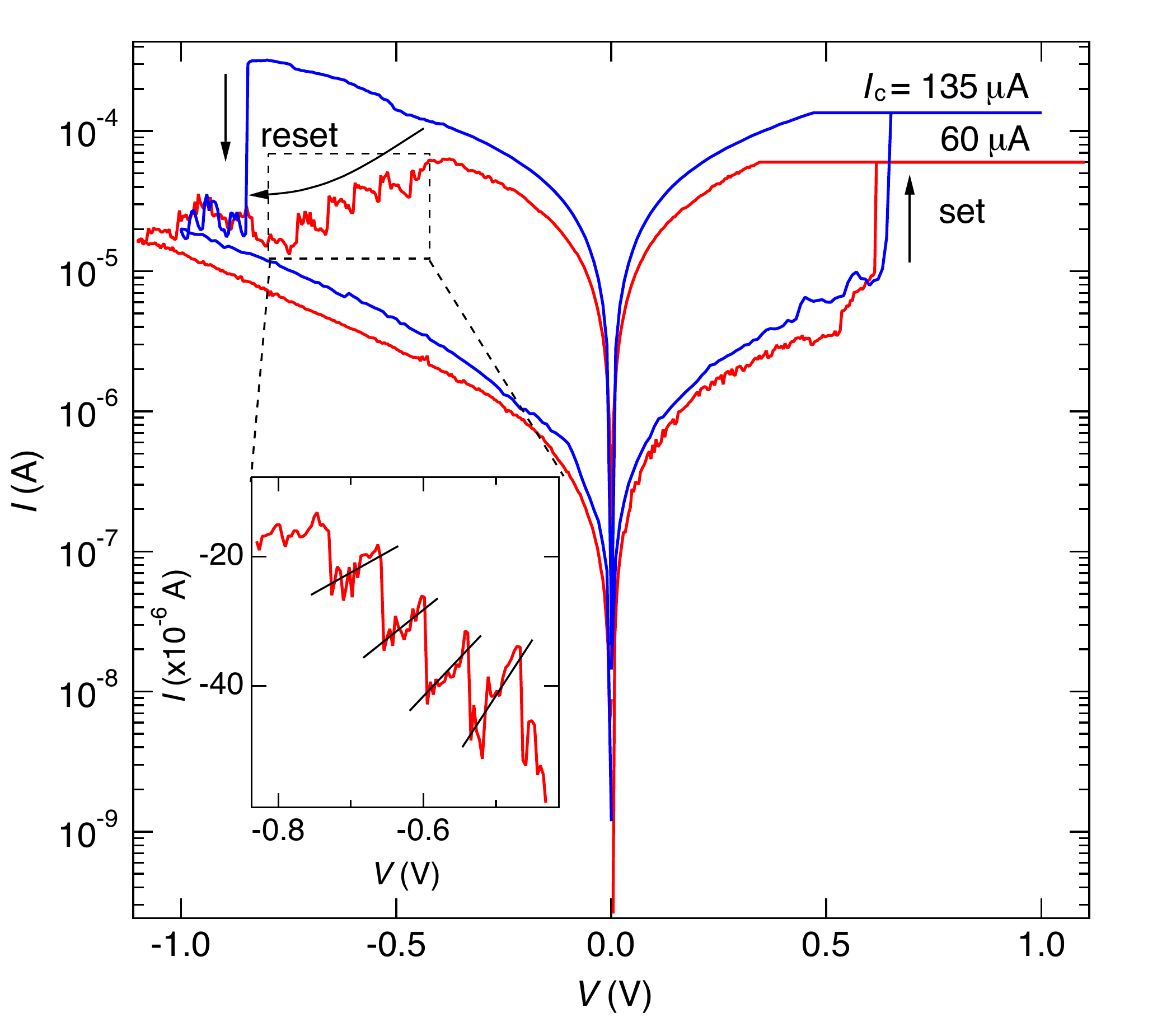}
	\caption{Currents in a log scale as functions of voltage with the compliance current for the set procedure $I_\mathrm{c}=135$ $\upmu$A (blue curve) and $60$ $\upmu$A (red curve), respectively. A stepwise feature is observed during the reset of the curve with $I_\mathrm{c}=60$ $\upmu$A, and its blown-up view in a linear scale is shown in the inset. The stepwise feature is absent from the curve with $I_\mathrm{c}=135$ $\upmu$A.}
	\label{main_iv}
\end{figure}

Electrical measurements are performed on several devices with the same structure as described in Section \ref{sec:device}, and the results are found to be similar and reproducible. The data presented in this paper are measured from a device with a cell area of 0.36 $\upmu$m$^2$. The forming procedure before the electrical measurements for each device is described in detail in Supporting Information Section \ref{sec:forming}. Two representative current-vs.-voltage ($I$--$V$) curves in the dc voltage sweep mode of the memristor operation are shown in Fig.~\ref{main_iv}, one with signatures of conductance quantization in the reset procedure and one without. A compliance current ($I_\mathrm{c}$) is applied during each set procedure to prevent permanent breakdown. The $I$--$V$ curve with $I_\mathrm{c}=135$ $\upmu$A (blue curve) shows the standard electrical characteristics with a steep drop in $I$ at $-0.85$ V in the reset procedure after the current reaches a maximum, whereas the one with $I_\mathrm{c}=60$ $\upmu$A (red curve) exhibits a series of descending steps (boxed with dashed lines) starting at a smaller bias voltage of $-0.38$ V. The steps correspond to the quantized conductance of a conducting channel in the switching layer, which reveals the nanoscale atomic-level reaction of a conductive filament consisting of oxygen vacancies in the switching layer \cite{Li2015,Lv2015,Long2013,Dirkmann2018,Capron2007}. During the reset procedure as the current is slowly switched from LRS to the high-resistance state (HRS), the quantum point contact of the filament that touches the negatively-charged top Ti electrode thins further and further because the oxygen vacancies of the filament are gradually removed under the voltage stress through recombination of the oxygen vacancies of the filament and oxygen ions from the Ti layer \cite{Dirkmann2018}. After the last oxygen vacancy in contact is removed, breaking the circuit established by the filament, a Schottky barrier is created in the conduction \cite{Syu2013}. 

\begin{figure}[t]%[h]
	\centering
\includegraphics[width=0.7\textwidth]{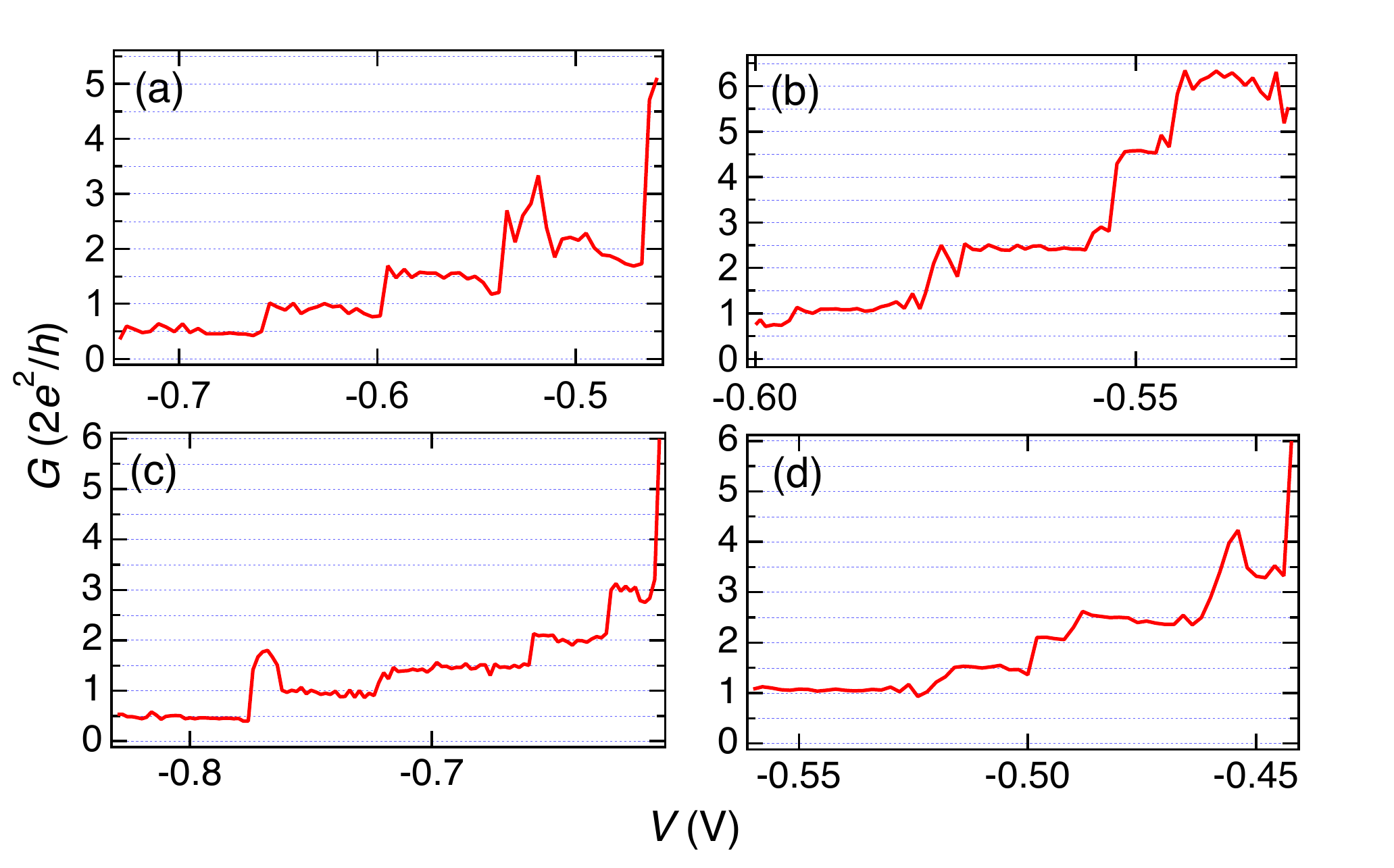}
	\caption{Representative examples of conductance quantization during the reset procedure with $I_\mathrm{c}=60$ $\upmu$A. Conductance plateaus occur at half-integer multiples of $2e^2/h$.}
	\label{G0}
\end{figure}

\begin{figure}[t]%[h]
	\centering
\includegraphics[width=0.7\textwidth]{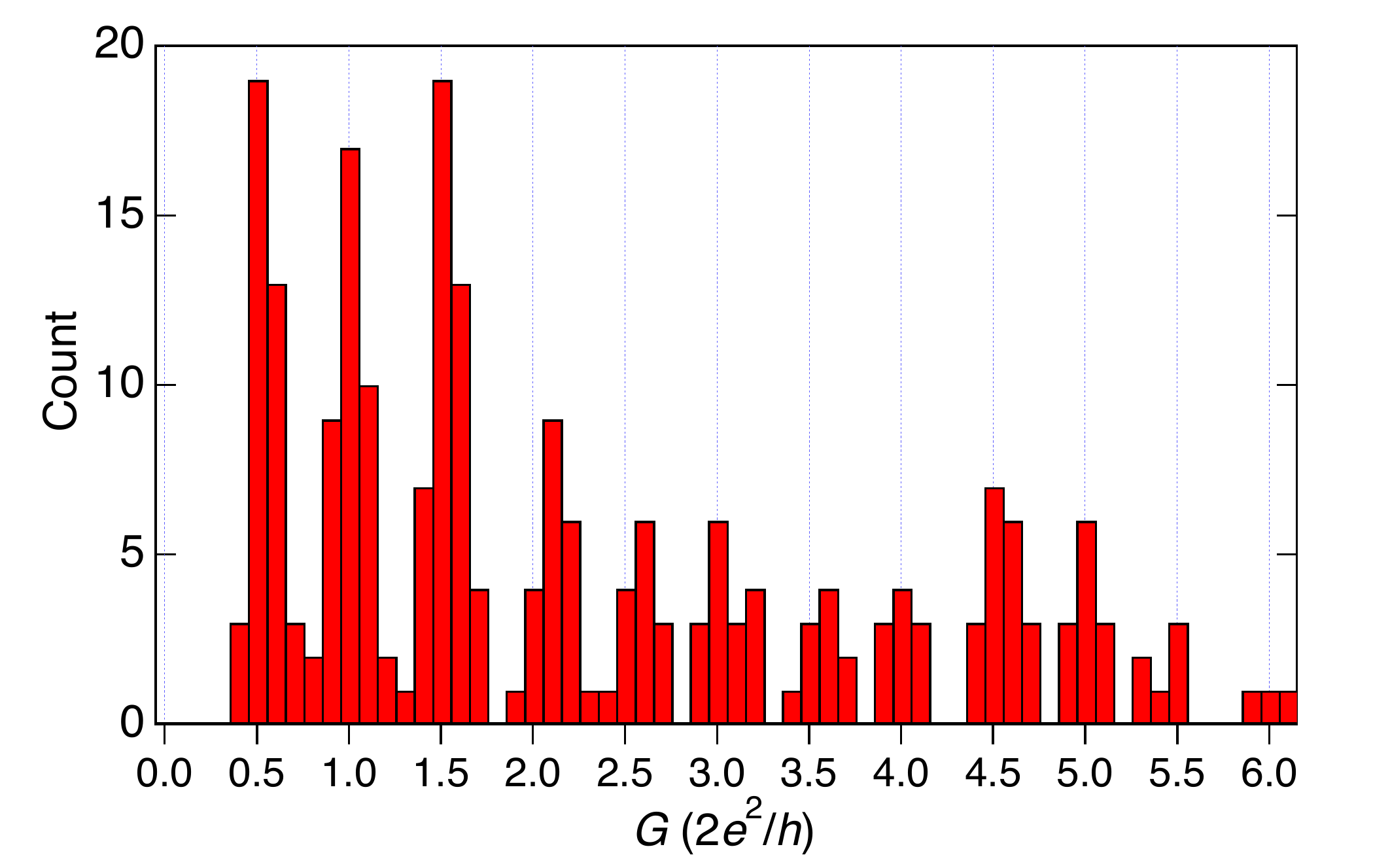}
	\caption{Histogram of the values of conductance plateaus retraced from 330 reset operations using the dc voltage sweep mode. More information about these 330 operations is listed in Table \ref{statistics}.}
	\label{count}
\end{figure}

The corresponding conductance of the quantum point contact in the reset procedure of the $I_\mathrm{c} = 60$ $\upmu$A curve in Fig.~\ref{main_iv} is calculated and expressed in terms of the conductance quantum $G_0 = 2e^2/h$ in Fig.~\ref{G0}(a), along with other examples of conductance quantization of the device shown in Figs.~\ref{G0}(b)--\ref{G0}(d). In each set-and-reset cycle, the voltage is swept at a rate of $\Delta V = \pm10$ mV per 0.5 second for each data point, except for the reset procedure from $-0.35$ V to $-1$ V , during which the sweep rate is decreased to $\Delta V = -2$ mV per 0.5 second to gently process the switching of the quantized conductance states. The resistance in series with the atomic point contact must be taken into consideration to precisely extract the quantized conductance of the point contact \cite{Lv2015,Tappertzhofen2015}. It can be seen that the conductance plateaus occur at some of the half-integer multiples of $G_0$. Fig.~\ref{count} summarizes the numbers of counts of respective values of the conductance plateaus collected from 330 $I$--$V$ curves using various $I_\mathrm{c}$  (listed in Table \ref{statistics}). The histogram clearly demonstrates the tendency of the device to yield quantized conductance. The appearance of half-integer multiples of $G_0$ instead of merely integer multiples is not universal in atomic point contacts. It is suggested to be caused by the chemical potential difference between the two carrier reservoirs across the filament \cite{Mehonic2013}, rearrangement of the atomic contact configuration \cite{Krishnan2017}, or possible weak magnetism from oxygen vacancies that may lift spin degeneracy \cite{Li2015}.

The compliance current $I_\mathrm{c}$ for the set procedure plays an important role in search of the quantized conductance states of a memristor. The set procedures have been performed with different $I_\mathrm{c}$ from 165 to 40 $\upmu$A, with 30 set-and-reset cycles completed for each $I_\mathrm{c}$. The statistics of the results from a memristor with a cell area of 0.36 $\upmu$m$^2$ are listed in Table \ref{statistics}. (Statistics from other devices with different cell areas show the similar behaviors; see Supporting Information Section \ref{sec:cell}. Temperature dependence of the electrical characteristics is also studied, as shown in Supporting Information Section \ref{sec:temp}.) \textit{All} the curves that exhibit conductance quantization in the reset procedure (see the row ``w/ Quant.") belong to the \textit{fair-set} group (definition in the next paragraph). The chance of observing conductance quantization in the reset procedure stays zero for $I_\mathrm{c} = 165$ $\upmu$A to 105 $\upmu$A, and then gradually increases to $50\%$ as $I_\mathrm{c}$ is gradually decreased to 70 $\upmu$A, and then reaches the maximum $67\%$ when $I_\mathrm{c}= 60$ $\upmu$A. The percentage then decreases to $33\%$ as $I_\mathrm{c}$ is decreased further to 40 $\upmu$A. This is an $I_\mathrm{c}$ so small that a \textit{good set} (definition in the next paragraph) can barely be acquired. With the highest yield of conductance quantization, $I_\mathrm{c}=60$ $\upmu$A is considered to be the optimal condition for later operation of the memristor for controlling the quantized conductance states.

\begin{table}[t]%[h]
\small
\centering
\setlength{\tabcolsep}{0.29em} 
{\renewcommand{\arraystretch}{1.1}
\begin{tabular}{lrrrrrrrrrrr}
\hline
$I_\mathrm{c}$ ($\upmu$A) & 165 & 150 & 135 & 120 & 105 & 90 & 80 & 70 & 60 & 50 & 40 \\ 
\hline
a) Good set                 & 30  & 30  & 30  & 30 & 30 & 23 & 20  & 11  & 5  & 4  & 0  \\ 
\rowcolor{Gray}
{\,\,\,\,\,\,\,w/ SCLC}      & {19}  & {18}  & {20}  & {17} & {20} & {11} & {7}  & {4}  & {3}  & {0}  & {0}  \\ 
b) Fair set                 & 0 & 0 & 0 & 0 & 0 & 7  & 10   & 18   & 21   & 15   & 17   \\ 
\rowcolor{Gray}
{\,\,\,\,\,\,\,w/ \textbf{Quant.}}      & {0}  & {0}  & {0} & {0} & {0} & {6} & {10}  & {15}  & {20}  & {12}  & {10}  \\ 
c) Poor set                 & 0  & 0  & 0  & 0  & 0  & 0  & 0   & 1   & 4   & 8   & 10   \\ 
d) Set failure               & 0  & 0  & 0  & 0  & 0  & 0  & 0   & 0   & 0   & 3   & 3   \\
\hline
\end{tabular}
}
\caption{Numbers of counts of various set conditions using different $I_\mathrm{c}$. (a) Good sets. Bottom row: good sets with SCLC in LRS. (b) Fair sets. Bottom row: fair sets with quantized conductance in the reset procedure. Notice that the fair-set category is the \textit{only} category in which current quantization in the reset procedure can be observed.  (c) Poor sets. (d) Set failures.}
\label{statistics}
\end{table}

\begin{figure}%[h]
	\centering
\includegraphics[width=0.7\textwidth]{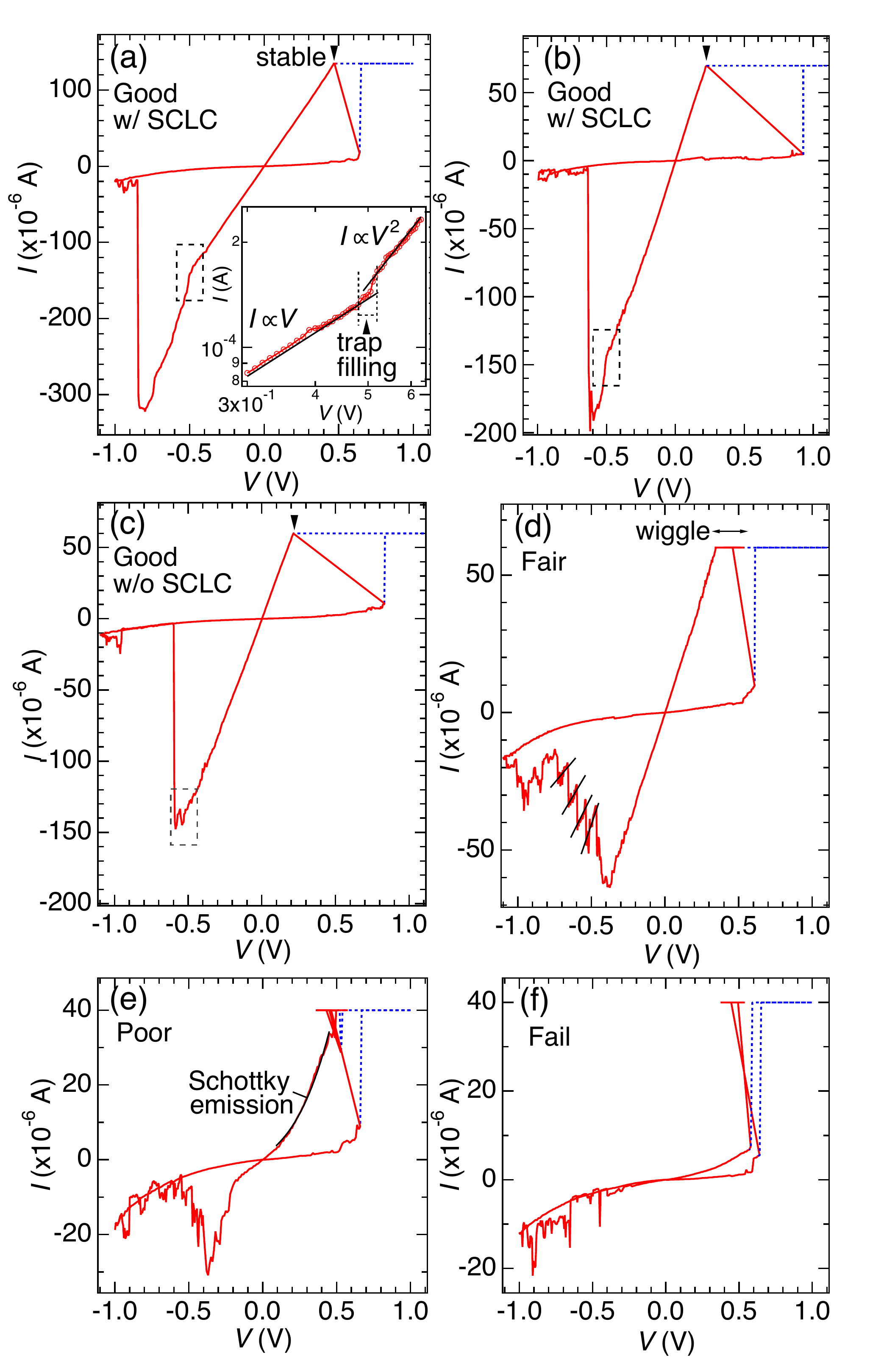}
	\caption{The electrical characteristics of the set-and-reset cycles can be classified into five categories: (a)(b) good set with SCLC, (c) good set without SCLC, (d) fair set, (e) poor set, and (f) set failure. The blue dashed curves are plotted against the programed bias voltage provided by the voltage source, whereas the red solid curves are plotted against the \textit{measured} bias voltage.}
	\label{niceset}
\end{figure}

The electrical characteristics of the set-and-reset cycles can be generally classified into five categories, as illustrated in Fig.~\ref{niceset} with representative examples. (More examples are presented in Supporting Information Section \ref{sec:set}.) The blue dashed curves are plotted against the programed bias voltage provided by the voltage source, whereas the red solid curves are plotted against the \textit{measured} bias voltage ($V_\mathrm{m}$). The only discrepancy between them lies in the set procedure when the resulting current abruptly jumps high to hit $I_\mathrm{c}$. The five categories are as follows: 

\begin{enumerate}
  \item \textbf{Good set with space-charge-limited current (SCLC)} (Figs.~\ref{niceset}(a) and \ref{niceset}(b)): The conduction in LRS after the set procedure is ohmic with a resistance of 2.0--4.0 k$\Omega$ (mostly around $3$ k$\Omega$), followed by a significant slope increase (boxed with dashed lines) at a negative voltage ($-0.48$ V in the representative example), known as the SCLC feature. The current then gradually increases to a maximum before it drops abruptly into HRS in a reset procedure (at $-$$0.85$ V in Fig.~\ref{niceset}(a) and $-$$0.61$ V in Fig.~\ref{niceset}(b)). A good set procedure is featured with an $I$--$V_\mathrm{m}$ curve hitting $I_\mathrm{c}$ at only one point for a stay, indicating a very stable $V_\mathrm{m}$.
  
  \item \textbf{Good set without SCLC} (Fig.~\ref{niceset}(c)): This has similar features with the previous category, except that it tends to undergo the reset process at smaller voltages ($\sim$0.18 V smaller \textit{on average}), and the SCLC signature is missing. (At the spot in the dashed box, it seems almost entering the SCLC regime especially when compared with Fig.~\ref{niceset}(b), but the filament fails to hold for it. More discussion about the SCLC is presented in the following contexts and in Supporting Information Section \ref{sec:goodset}.) Tiny fluctuations are observed at larger negative bias voltages in LRS, before the reset procedure takes place (at $-$$0.59$ V in this example).
  
  \item \textbf{Fair set} (Fig.~\ref{niceset}(d)): The conduction in LRS after the set is ohmic with a resistance of 4.8--7.5 k$\Omega$ (mostly around $6$ k$\Omega$), which is about twice larger than those in the good-set cases. Tiny fluctuations are usually observed at larger bias voltages in the ohmic state. The reset procedure starts at an even smaller bias voltage ($-$$0.38$ V in this example), but with a progressive phase that is prolonged to a more negative bias voltage. A fair set procedure is featured with an $I$--$V_\mathrm{m}$ trace frequently wiggling left and right along the horizontal compliance line. This is the \textit{only} category in which current quantization in the reset procedure can be observed. Notice that $I_\mathrm{c}=60$ $\upmu$A in both Figs.~\ref{niceset}(c)  and \ref{niceset}(d). In very few cases where quantization is absent (not presented in Fig.~\ref{niceset}), the reset process exhibits chaotic noise-like fluctuations similar to those found in Fig.~\ref{niceset}(e).
  
  \item \textbf{Poor set} (Fig.~\ref{niceset}(e)): The conduction in LRS is no longer ohmic, but follows the Schottky-emission equation, with resistance considerably higher than that in the fair-set cases. A poor-set procedure is also featured with an $I$--$V_\mathrm{m}$ trace wiggling along the compliance ceiling, and even occasionally falling off and rising back to the ceiling before entering LRS.
  
  \item \textbf{Set failure} (Fig.~\ref{niceset}(f)): The $I$--$V$ curve cannot enter LRS after the set procedure.
\end{enumerate} 

The Schottky-emission equation is shown as follows:
\[ {I}=aA^*{T}^{2} \exp \left[\frac{-(\phi_\mathrm{B}-\sqrt{{q^3 V}/{4 \pi \epsilon d_\mathrm{s}}})}{k T}\right], \]
where $a$ is the effective cross section of the filament, $A^*$ is the effective Richardson constant, $q$ is the carrier charge, $T$ is the temperature, $\phi_\mathrm{B}$ is the energy barrier height, $d_\mathrm{s}$ is the effective switching thickness, and $\epsilon$ is the permittivity, which is $\sim$$25 \epsilon_0$ for HfO$_2$. A fit to the LRS curve in Fig.~\ref{niceset}(e) yields $\phi_\mathrm{B} = 0.36$ eV and $d_\mathrm{s}=2.4$ nm, which reveals the characteristics of the device structure with the filament grown between the bottom TiN electrode and the positively-biased top Ti electrode to a point where it is only a tiny gap ($\sim$$2.4$ nm) away from completion of the connection. The value of $\phi_\mathrm{B}$ is about half of that of our previous similar HfO$_x$ memristors \cite{Syu2013}, which may imply a larger amount of impurities or defects in the switching layer.

Shown in the inset of Fig.~\ref{niceset}(a) is a log-scale blown-up view of the spot where the $I$--$V$ slope changes due to SCLC, which is known to be mostly detected in memristors with $\phi_\mathrm{B}\lesssim 0.3$ eV \cite{Malliaras1999}. The $I$--$V$ characteristic is ohmic until $V$ is swept to a more negative value of $-0.48$ V, where the device enters the trap-filling regime \cite{Kao1981,Lampert1970,Mark1962} with the slope in the log scale prominently increased to $> 2$, until the $I$--$V$ characteristic changes again to follow the Mott--Gurney law of SCLC \cite{Mott1940} starting at $V=-0.52$ V: 
\[ I=\frac{9}{8} a\epsilon \mu \frac{V^{2}}{d_{s}^{3}}, \]
where $\mu$ is the electron mobility, which is $\sim$200 cm$^2$/Vs in HfO$_2$ \cite{Negara2009,Ozben2010}. The $V^2$ law of SCLC only holds for a limited range of bias voltage. After $V$ is swept to $-0.64$ V, the characteristic becomes $I\propto V^x$ with $1< x <2$, until the reset procedure starts. This may be interpreted with a negative field dependence of the mobility of the space charges due to positional and energetic disorder in the material \cite{Bassler1993,Bhattarai2018}. At lower electric fields, the most energetically favorable paths for percolative hopping transport will proceed via randomly oriented jumps. However, with increasing electric field, charge carriers are forced to make less energetically favorable jumps in the direction of the field, leading to a reduced mobility. 

It is generally believed that a larger $I_\mathrm{c}$ leads to a more compact structure of the oxygen vacancies \cite{Zahoor2020} or a larger diameter \cite{Yu2016} of the conductive filament. Therefore, a large $I_\mathrm{c}$ in our experiment such as 135 $\upmu$A results in an $I$--$V$ curve with standard characteristics of a fairly strong filament, as shown in Fig.~\ref{niceset}(a). For $I_\mathrm{c}=60$ $\upmu$A, however, manifolds of set conditions and electrical characteristics of the cycles can be observed (see Table \ref{statistics}), despite it being the set $I_\mathrm{c}$ with the highest yield of conductance quantization in the reset procedure. Figs.~\ref{niceset}(c) and \ref{niceset}(d) both have $I_\mathrm{c}=60$ $\upmu$A, and the key difference between them is the stability of $V_\mathrm{m}$ in the set procedure. One may forecast conductance quantization in the later reset procedure only when a ``wiggling", unsteady $V_\mathrm{m}$ is detected in the set procedure. 

Some previous works have tried to determine the size of the conductive filaments in Ti/HfO$_2$/TiN memristors either through TEM material analyses \cite{Privitera2013} or through theoretical simulations \cite{Dirkmann2018,Nardi2012}. From the electrical characteristics and the TEM images or simulation results provided in these works, a filament in the 10-nm thick HfO$_2$ layer of our device with a resistance of around 3 k$\Omega$ to 6 k$\Omega$ may be roughly estimated to be only $\lesssim 3$ nm in diameter. This explains why it has been extremely difficult to observe a filament in cross-sectional TEM images of our devices.  Electrical stresses on a narrow filament structure that lead to conductance evolution in the order of $G_0$, on the other hand, have also been studied in several simulation works \cite{Li2015,Li2017,Long2013,Zhong2016}. However,  a precise prediction of the quantized conductance value as a function of the atomic evolution of the filament structure is not available yet, nor is there a concise conclusion on the numerical values of the stress voltage to optimize the chance of observing conductance quantization. More experimental and theoretical research are necessary to unveil the detailed mechanism of the conductance quantization, and our work presents a step forward toward understanding the filament evolution.

Although multiple growths of filaments or branches composed of oxygen vacancies are possible in the device, the conduction is believed to be contributed by a single dominant filament because there is only one LRS in each $I$--$V$ cycle (i.e., multiple filaments would have been resulted from multiple set procedures in sequence in an $I$--$V$ cycle, exhibiting state switching between multiple LRSs). Once a filament is established, the current flows mostly through the connected filament, and further filament growth will be suppressed owing to reduced electric field \cite{Kwon2010}. It has been found from the TEM images of SiO$_2$-based planar devices that, in the LRS, there exists only one completed filament accompanied with a few incomplete ones \cite{Yang2012}.

A good set without SCLC (Fig.~\ref{niceset}(c)) may be regarded as an intermediate state between the state with SCLC (Figs.~\ref{niceset}(a) and \ref{niceset}(b)) and the state with conductance quantization (Fig.~\ref{niceset}(d)) in the sense of the resultant filament strength. For a filament robust enough to stand a higher negative voltage, the device can enter the SCLC-dominating regime until an abrupt drop of the current occurs upon the reset procedure when the filament is ruptured under a much higher voltage. A filament that exhibits a progressive reset, on the other hand, features a relatively unstable figure showing unsteady $V_\mathrm{m}$ in the set procedure, possibly with the oxygen vacancies at the tip of the filament moving around among multiple metastable states to establish or dismiss an ohmic contact. This instability is revisited in the reset procedure starting around $V = -0.4$ V in a more distinct fashion, i.e., the conductance quantization, which again reveals nanoscale atomic-level movements. As $I_\mathrm{c}$ is lowered to 40 $\upmu$A, with examples shown in Figs.~\ref{niceset}(e) and \ref{niceset}(f), 43\% of the $I$--$V$ curves exhibit Schottky emissions or set failures. From the facts above, $I_\mathrm{c}=60$ $\upmu$A is the critical compliance current in our experiment that is just large enough to build an ohmic filament, and yet simultaneously small enough to allow atomic-level behaviors to be unveiled in the memristor operation. Our experiment is the first of its kind to classify the different signatures of $I$--$V_\mathrm{m}$ (the \textit{measured} voltage) for a better understanding of the atomic-level dynamics in a memristor. 

Electrochemical metallization memristors may behave differently from VCMs in the reset procedure in accordance with the filament resistance. For example, Celano \textit{et al.}~\cite{Celano2015} have found from Cu/Al$_2$O$_3$/TiN memristors that Cu filaments with larger diameters and lower LRS resistance tend to exhibit progressive resets, whereas Cu filaments with smaller diameters and higher LRS resistance are inclined to undergo abrupt ruptures, which is opposite to our findings on the HfO$_x$-based devices. The different behaviors can be interpreted with the filament growth and dissolution dynamic scenario being governed by the ion or vacancy mobility and diffusivity, and the redox rate \cite{Yang2013a, Ielmini2016}. The activation energies for diffusion of oxygen vacancies in HfO$_2$ ($\sim$0.7 eV for a mobile charged (2+) oxygen vacancy \cite{Capron2007} and $\sim$3 eV for a neutral one \cite{Duncan2016}) are much higher than those of Cu ions in amorphous Al$_2$O$_3$ ($\sim$0.3 eV for mobile ones \cite{Pandey2018} and $\sim$0.9 eV for less mobile ones \cite{Sankaran2012}). As a very narrow oxygen-deficient filament undergoes a reset procedure with extremely limited current and thus limited Joule power, the oxygen vacancies, which have significantly low diffusivity, may leave the filament one by one slowly and discretely, allowing us to observe the step-wise conductance. The tendency to exhibit progressive resets at lower $I_\mathrm{c}$ and consequently with higher LRS resistance is typical of memristors based on the VCM mechanism \cite{Wouters2019,Volos_book}.

It is not clear yet why set-and-reset cycles with the same $I_\mathrm{c}$ (60 $\upmu$A) can lead to different set conditions for the same memristor. Possible overshooting of the current is considered at the compliance point as the memristor is quickly switched from HRS to LRS. The $I$--$V$ curves of our device exhibit a fairly linear relation in LRS, as shown in Figs.~\ref{niceset}(c) and \ref{niceset}(d). This ohmic behavior indicates that possible parasitic capacitance, and hence current overshooting, are quite limited in our device \cite{Ambrogio2016}. However, because the conductance characteristics of the memristor are markedly affected by the dynamic behaviors in the nanoscale, possible small overshooting of the current even to a minimal extent may matter. Since it is difficult to directly detect the probable variations of this minimal overshooting, monitoring $V_\mathrm{m}$ becomes the only practical and effective method to determine the set condition right away. The cause of the different resultant set conditions under the same $I_\mathrm{c}$ may also involve the instant internal state of the memristor in the atomic scale being affected dynamically by the second-order state variables (the temperature decay for example) present in the structure \cite{Kim2015}. With these nanoscale uncertainties in the system, it seems that the most easily observable signal that reveals the multiple metastable states of a point contact in the filament, and thus the potential to yield quantized conductance states in the reset procedure, is the wiggling $V_\mathrm{m}$ at the set procedure.  Monitoring $V_\mathrm{m}$ therefore becomes the critical method to track the qualities of the device fabrication and measurements.

\subsection{Measurements with the Pulse-Mode Reset Procedure}\label{sec:pulse}

\begin{figure}[h]
	\centering
\includegraphics[width=0.75\textwidth]{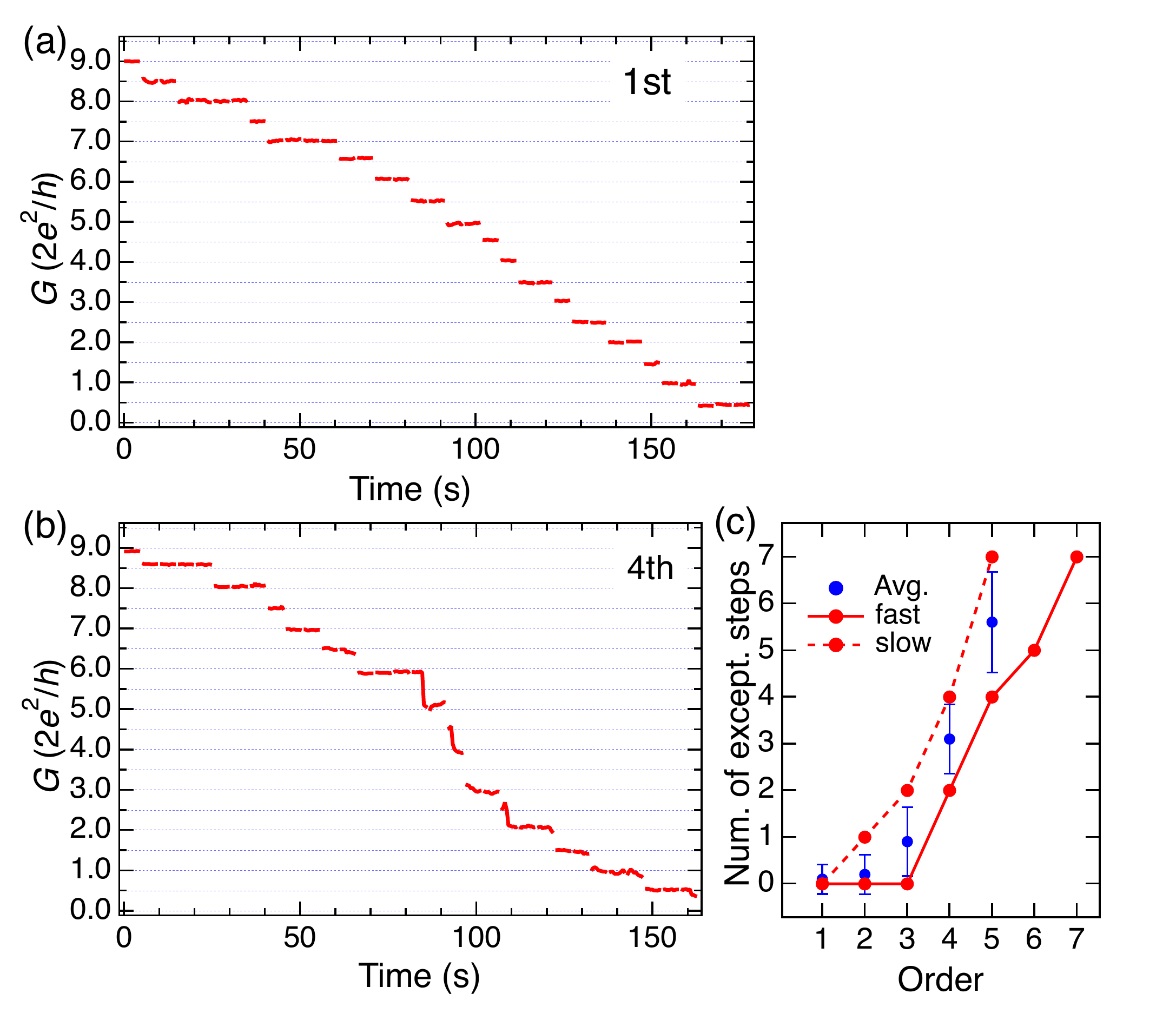}
	\caption{(a)(b) Two representative examples of controllable quantized conductance states at integer multiples of $0.5G_0$ using a pulse-mode reset procedure after a \textit{fair set} with $I_\mathrm{c}=60$ $\upmu$A. (a) is taken right after changing $I_\mathrm{c}$ from 70 $\upmu$A to 60 $\upmu$A. (b) is taken after three successive cycles with fair sets. (c) Average total number of exceptional steps, which is the sum of spontaneous steps with $|\Delta G| \leq G_0$ and stimulated steps with $|\Delta G| = G_0$, as a function of the order of successive fair-set cycles. Also displayed are the fastest and the slowest training sequences from the data.}
	\label{pulse2}
\end{figure}

Distinguishing \textit{fair sets} from the others allows us to achieve a high success rate of control of the conductance quantization of the memristor using a pulse-mode reset procedure. A typical example is demonstrated in Fig.~\ref{pulse2}(a). The reset process is preceded by a set procedure using $I_\mathrm{c}=60$ $\upmu$A that exhibits a fair-set condition (i.e., with wiggling $V_\mathrm{m}$) as depicted in Fig.~\ref{niceset}(d). Voltage pulses with fixed width of 0.1 second and fixed value of $-0.35$ V are used to control the atom-by-atom evolution. The pulse width and amplitude are chosen from amongst multiple tests to achieve the optimal result, that is, to stimulate switching to the next conductance state with a minimal average number of pulses. The pulse value $-0.35$ V, which is very close to the onset voltage of the reset process observed in dc voltage sweeps with a fair set (Fig.~\ref{niceset}(d)), is speculated to be a favorable value for our device to activate recombination between oxygen ions and vacancies through oxygen migration by providing a proper electric field and a local temperature enhancement due to Joule heating \cite{Dirkmann2018,Roldan2021}. After each pulse, the current is read at $-0.01$ V for 5 seconds, from which the conductance of the point contact is computed and then presented in units of $G_0$. It can be seen that the conductance decreases stepwise from $9G_0$ to $0.5G_0$ in steps of $0.5G_0$ with great precision, with an average standard deviation of only $\sim$$0.014G_0$ for the quantized plateaus. The width of each conductance plateau falls within 20 seconds, which corresponds to 1 to 4 voltage stimuli before stepping down to the next plateau.

Fig.~\ref{pulse2}(b) presents another example of the pulse-mode measurement of the same memristor after a fair set using the same $I_\mathrm{c}$ ($60$ $\upmu$A). This set of data is from a cycle taken after three successive cycles with fair sets using $I_\mathrm{c} = 60$ $\upmu$A. (The quantized conductance data of all the successive cycles of this sequence are presented in Supporting Information Section \ref{sec:training}.) In this 4th cycle, not every step in the reset procedure is $0.5G_0$ in height and stimulated by a voltage pulse. A few exceptions are found between $6G_0$ and $2G_0$, where some of the conductance drops have magnitudes of $1G_0$, and some occur spontaneously without a pulsed voltage stimulus. This leads to a faster switching from a high-conductance state to the lowest one. The occurrence probability of these exceptions roughly increases with the number of repetitions of the measurements under the same $I_\mathrm{c}$ ($60$ $\upmu$A), implying a possible training effect. This training effect can be removed by applying a higher $I_\mathrm{c}$ ($70$ $\upmu$A for example) to the set procedure before returning to the original $I_\mathrm{c}$ to regain well-controlled state switchings in steps of $0.5G_0$ without exceptions. The training effect may be seen from the statistics shown in Fig.~\ref{pulse2}(c), where the average number of exceptions (i.e., the number of  spontaneous steps with $|\Delta G| =0.5G_0$ or  $G_0$, plus the number of  stimulated steps with $|\Delta G| = G_0$) is plotted against the order of successive fair-set cycles, collected from 10 sequences in which no steps larger than $G_0$ are involved. (In other words, a very minor group of sequences interrupted by a reset procedure with steps larger than $G_0$ are not included in Fig.~\ref{pulse2}(c) for simplicity.) For example, the total number of this kind of exceptional steps is 4 for Fig.~\ref{pulse2}(b). Also displayed in Fig.~\ref{pulse2}(c) are the fastest (dashed line) and the slowest (solid line) training sequences from the data. It is rare for the fair set to appear continuously for more than 5 cycles, except for a ``slow learner" like the solid trace in Fig.~\ref{pulse2}(c), which has 7 successive fair sets. Some may hope for an ideal memristor with controllable quantized conduction that always generates a conductance decrease of $0.5$$G_0$ in response to each voltage stimulus, but no realization of such a memristor has been reported up to date. It is possible that the conductance is governed not only by the external stimuli but also by its instant internal state, known as the property of a second-order memristor. In fact, memristors with this kind of instability on intermediate conductance states are being proposed as candidate neuromorphic computing devices that can naturally emulate the temporal behaviors, including sequence learning, of biological synapses \cite{Mikheev2019,Marrone2019}. More research is necessary to accommodate or even take advantage of the second-order behaviors of the memristors for constructing practical neuromorphic computing architectures.

Set conditions other than fair sets generally do not yield controllable quantized conductance states in the pulse-mode reset procedure. For example, most of the time a reset process preceded by a good set requires a larger stimulating voltage, but only to lead to an abrupt conductance drop that brings the device directly to HRS. In very few cases, a reset process preceded by a good set that enters LRS at a relatively small $V_\mathrm{m}$ (similar to that in Fig.~\ref{niceset}(c)) can exhibit a few quantized conductance plateaus, but is far away from accessing a complete set of integer multiples of $0.5$$G_0$. Therefore, to efficiently repeat the operation of the quantum-level manipulation, a pulse-mode reset procedure is executed only when a fair set is detected. The ability of the nanoscale atomic structure of a filament to switch among multiple metastable states upon the set process, as implied by the wiggling $V_\mathrm{m}$ during a fair set (Fig.~\ref{niceset}(d)), may be a necessary feature for a memristor to permit excellent realization and modulability of the quantized conductance states. Future device fabrications and characterizations are encouraged to incorporate $V_\mathrm{m}$ measurements to analyze the set condition. Memristors that guarantee fair sets under certain $I_\mathrm{c}$'s should be favorable.

For comparison, Xue \textit{et al.}~\cite{Xue2020} have found from a Pt/HfO$_x$/ITO memristor that switching of the quantized conductance states needs to be stimulated by extremely long (20-second) pulses, and becomes even more insensitive to voltage stimuli at lower conductance, which they attributed to the lower current and thus a lower  power available for modulating the filament. In contrast to their findings, our devices are more efficient in that they are sensitive to short voltage stimuli throughout the whole reset procedure. This indicates the high modulability of a very narrow filament even with a very low current, which points to the criticality of the current density and the local temperature enhancement based on heat transfer around the constriction (i.e., the narrowest point) of the filament during the reset procedure \cite{Dirkmann2018,Roldan2021}. On the other hand, there are other previous studies on conductance quantization of memristors that also demonstrate state switching upon short pulsed voltage stimuli, but with much lower precision of the quantized conductance values \cite{Li2015}. As analytical differentiation of the $I$--$V_\mathrm{m}$ characteristics (Fig.~\ref{niceset}) is employed in our experiment during device selection and measurements, the precision and signal-to-noise ratios of the quantized conductance are significantly improved in our experiment compared to previous reports, and thereby brings the study of memristors closer to practical application in neuromorphic computing.

\section{Conclusions}

In summary, we report on controllable quantized conductance states of TiN/Ti/HfO$_x$/TiN memristors in a pulse-mode reset procedure with significantly improved precision. The high controllability and precision are realized through analytical diagnoses of the set conditions of the fabricated devices. The $I$--$V_\mathrm{m}$ characteristics of the set procedure can be classified into \textit{good}, \textit{fair}, and \textit{poor} conditions, and only those with fair sets (i.e., with a ``wiggling", unstable measured voltage $V_\mathrm{m}$ at the compliance current) can permit quantized conductance states in the reset procedure. Controlled conductance decrease from $9G_0$ to $0.5G_0$ in steps of $0.5G_0$ is successfully observed in pulse-mode reset procedures that are preceded by a fair set with an optimal compliance current ($60$ $\upmu$A). A training effect that leads to a faster state switching is found in the operation, which is regarded as a candidate mechanism for temporal sequence learning. Our experiment is the first of its kind to point out the importance of monitoring the measured bias voltage to track the qualities of the device fabrication and measurements for the research of conductance quantization of memristors. Our study unveils a full spectrum of the dynamic behaviors under different set conditions to provide an overview of the mechanisms of the conductive filament, from a strong ohmic structure with space-charge-limited current (SCLC), to that without SCLC, then to a relatively unstable configuration that supports quantized conduction as well as ohmic conduction, and then to a nano-gapped channel with Schottky emission. This allows a better understanding of the dynamics of the nanoscale atomic-level structures in the memristors, which should promote the progress of future design and fabrication of the memristors for neuromorphic computing and quantum information processing algorithm.

\section*{Associated Content}

Supporting Information available: more examples for each set category; information pertaining to the forming procedure, SCLC statistics, and operation using a pulse-mode set procedure; measurements of devices with different cell areas; temperature-dependent measurements down to 4.2 K; data of a complete training sequence.

\section*{Acknowledgements}
We acknowledge the groups of Prof.~Shu-Fen Hu and Prof.~Yann-Wen Lan for assisting us with the measurements. We thank Dr.~Chih-Yang Lin for helpful discussion. This study is sponsored by the Ministry of Science and Technology of Taiwan under Grant No.~MOST 109-2112-M-003-009 and MOST 110-2112-M-003-019.

\bibliography{_atomic_9-acs_c}
\newpage

\makeatletter\@input{_atomic_9-acs_c_aux.tex}\makeatother
\end{document}

\maketitle

All the $I$--$V$ figures in the Supporting Information show the device current plotted against the \textit{measured} bias voltage ($V_\mathrm{m}$).

\section{The Forming Procedure}\label{sec:forming}

\begin{figure}[!htb]
\centering
\includegraphics[width=0.8\textwidth]{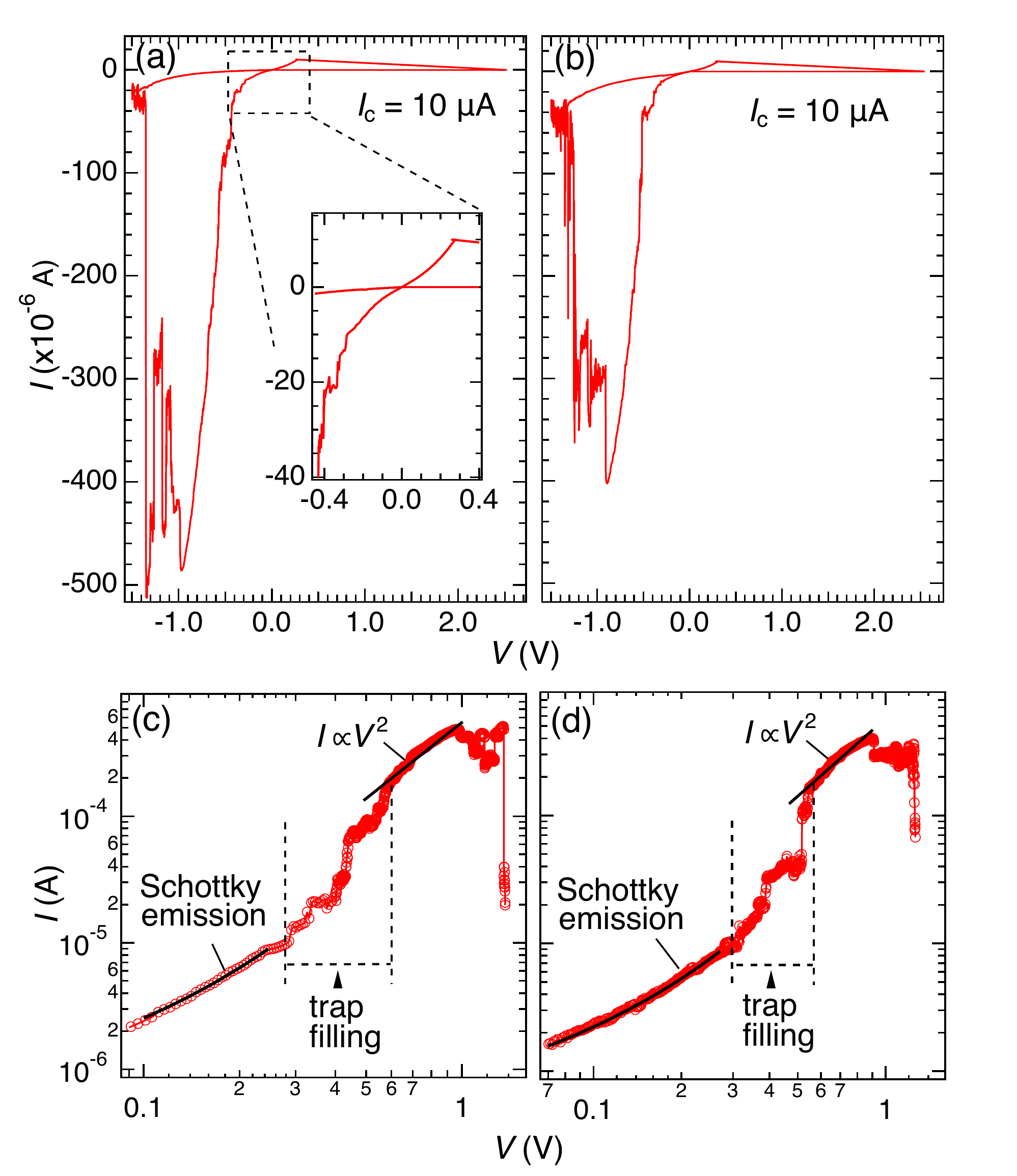}
\caption{Representative $I$--$V$ curves of forming procedures for devices with a cell area of (a) 0.36 $\upmu$m$^2$ and (b) 0.60 $\upmu$m$^2$, respectively. (c) and (d) are the log-scale view of the negative-voltage side of (a) and (b), respectively. }
\label{forming}
\end{figure}

The forming procedure for all the devices are performed in the dc voltage sweep mode with a compliance current $I_\mathrm{c}$ ranging from 10 to 100 $\upmu$A. The bias voltage for the forming procedure is programmed to be swept from 0 to 3.5 V and then back to 0 V at a rate of $\Delta V = \pm2$ mV per 0.5 second for each data point. $I_\mathrm{c}$ starts from 10 $\upmu$A, and will be increased by 10 $\upmu$A each time up to 100 $\upmu$A only if the forming procedure is not successful under the present smaller $I_\mathrm{c}$. Devices that cannot undergo a successful forming procedure with an $I_\mathrm{c} \leq 100$ $\upmu$A are considered defective.

Representative examples of a successful forming procedure are shown in Fig.~\ref{forming}. Most devices can undergo a successful forming process with $I_\mathrm{c}$ = 10 $\upmu$A, followed by a Schottky-emission-like LRS, and then a tremendous increase of the conductance starting around $-0.3$ V. The current keeps enhancing enormously and becomes comparable with that from the category “good set with SCLC” around $-0.7$ V. Under a higher negative voltage, the $I$–$V$ curve starts to experience some violent oscillations before it enters HRS in a reset procedure. 

The tremendous increase in the conductance under negative voltage is also examined in the log scale, as plotted in Figs.~\ref{forming}(c) and \ref{forming}(d) for Figs.~\ref{forming}(a) and \ref{forming}(b), respectively. It can be seen that the $I$--$V$ curve shows a Schottky-emission behavior (both with $\phi_\mathrm{B} \approx 0.36$ eV and $d_\mathrm{s} \approx 1.9$ nm) at lower voltages, then a tremendous increase in $I$ with significant oscillations, and then an $I$--$V$ relation close to the $V^2$ law of SCLC under higher voltages. The oscillating but overall increasing current in the middle regime is suspected to be attributed to trap filling. It is presumed that a local trap-rich conducting region is formed in the HfO$_2$ layer after the forming process \cite{Voronkovskii2019}, and the new-born filament which initially exhibited a Schottky-emission behavior can be unsteady and susceptible when handling the trap-mediated SCLC for the first time, thereby leading to apparent oscillations.

The non-ohmic $I$--$V$ in the LRS right after the forming process does not indicate a device of poorer quality. In fact, in a study on HfO$_2$-based memristors by Milo \textit{et al.}~\cite{Milo2019}, the memristor with the best quality (i.e., with  the highest process yield,  lowest device-to-device variability, lowest set and reset voltages, largest resistance ratio, and best endurance) happens to be an Al-doped HfO$_2$ device with a curved, non-ohmic $I$--$V$ in the LRS after the forming process. In our device, the SCLC-dominating high conduction through the newly-formed filament under higher negative voltages before the 1st reset has probably invoked some modulation of the filament configuration (which may have led to the oscillations in the trap-filling regime) into a stronger one, and thereby the filament remains after rupture help the growth of an ohmic filament in the following set procedures.

Regarding the 1st reset procedure after forming, there may be step-like features in those violent oscillations during the reset, but the noise levels of possible conductance plateaus among them are greater than $G_0$, and thus their corresponding values in conductance quanta are  unreliable. Attempts on observing conductance quantization using a pulsed-voltage mode during the 1st reset procedure were also not successful.

\section{The Set Procedure}\label{sec:set}
This section provides more examples for each set category, and discussion on the space-charge limited current (SCLC). All the data are from the same device as depicted in the paper (cell area = 0.36 $\upmu$m$^2$).

\subsection{Good Set}\label{sec:goodset}

\begin{figure}[!htb]
\centering
\includegraphics[width=1.15\textwidth,center]{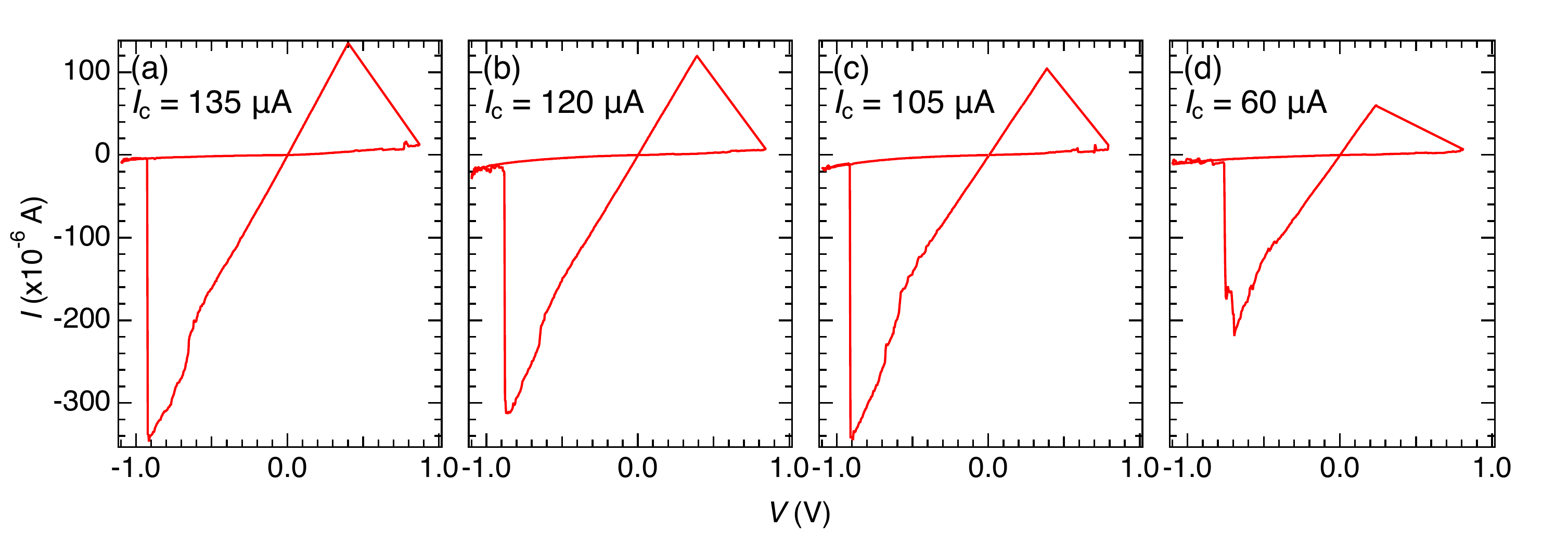}
\caption{Representative $I$--$V$ curves of good sets \textit{with} SCLC.}
\label{goodwSCLC}
\end{figure}

\begin{figure}[!htb]
\centering
\includegraphics[width=1.15\textwidth,center]{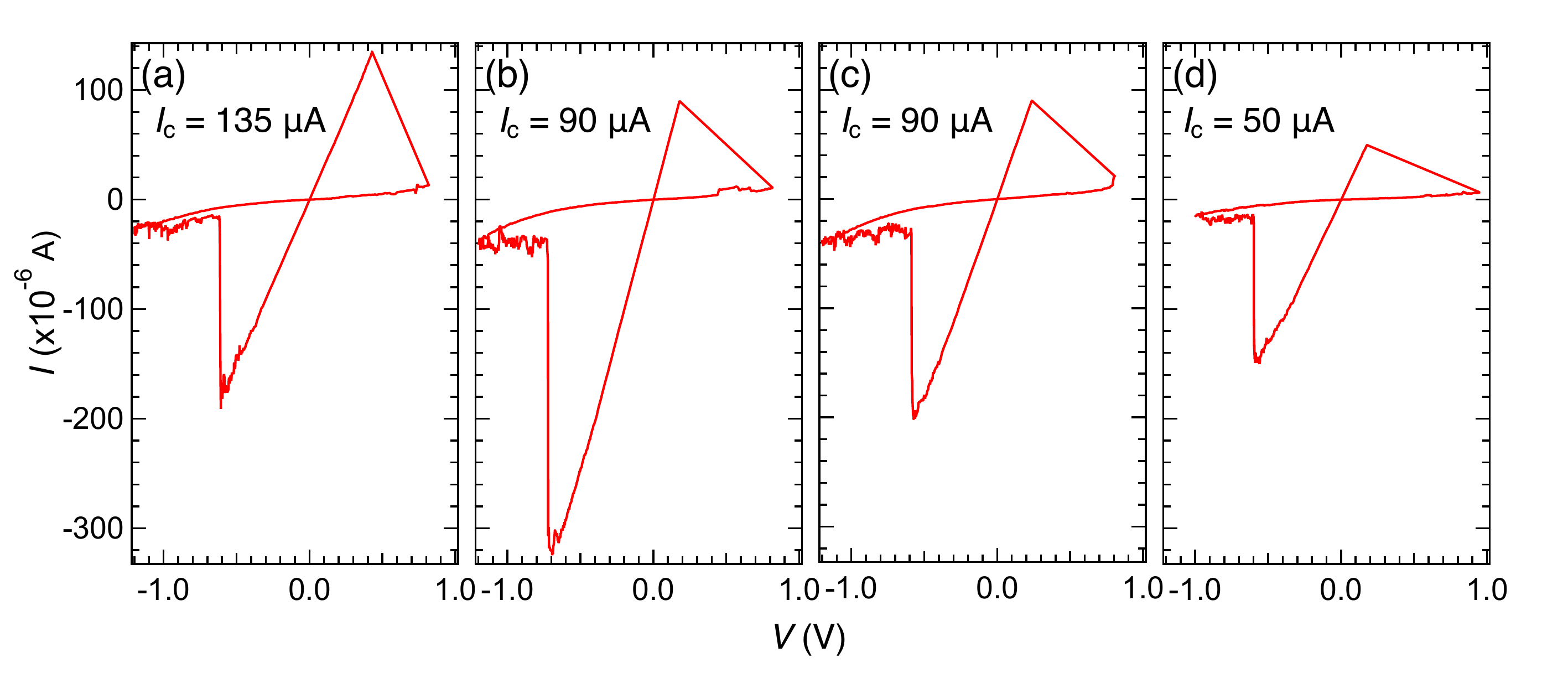}
\caption{Representative $I$--$V$ curves of good sets \textit{without} SCLC.}
\label{goodwoSCLC}
\vspace{0.8cm}
\end{figure}

\begin{figure}[!htb]
\centering
\includegraphics[width=0.62\textwidth,center]{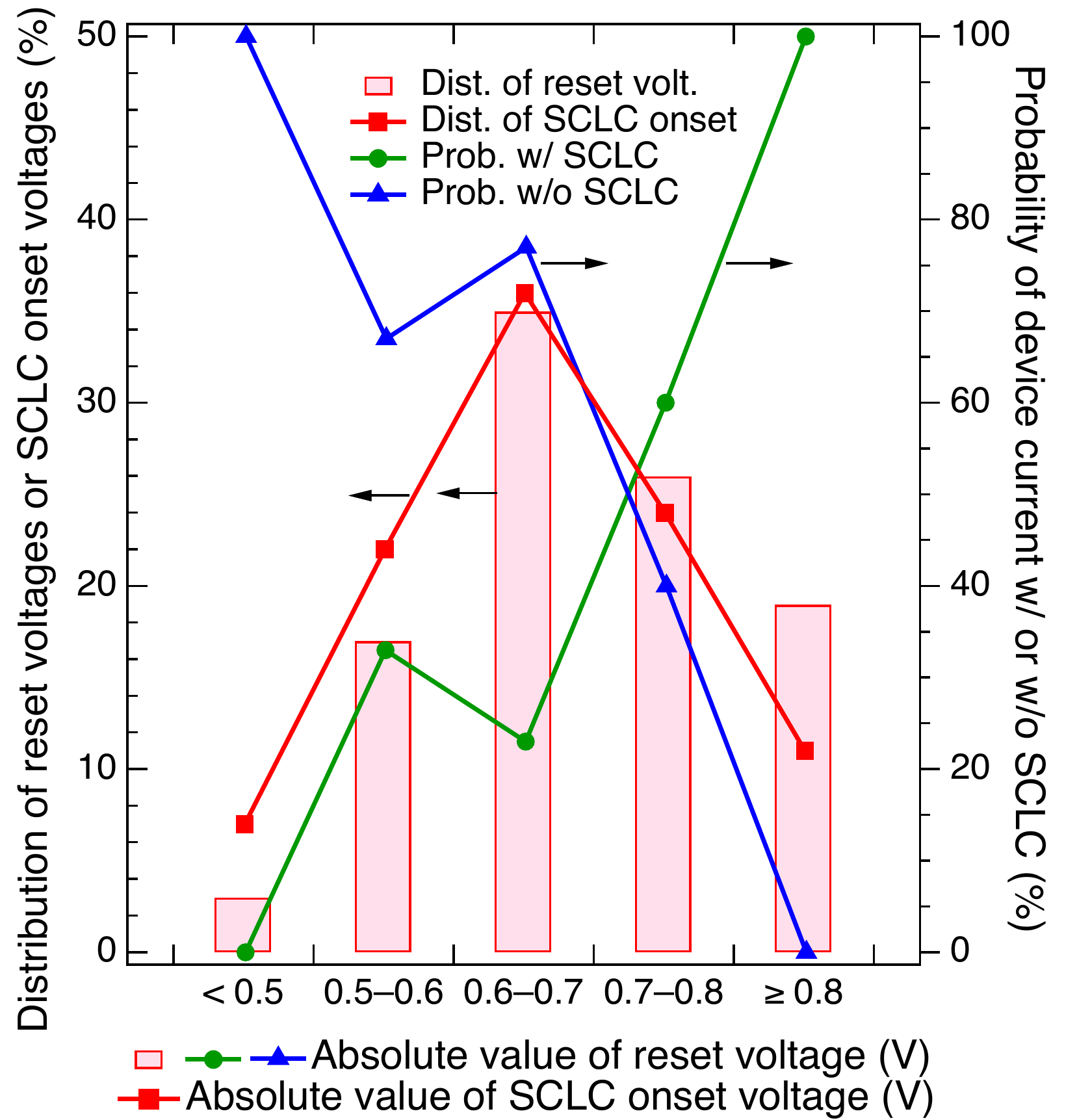}
\caption{Distributions of the reset voltages and the SCLC onset voltages, and the probability of observing or not observing the SCLC feature in an $I$--$V$ curve with the respective reset voltage.}
\label{dist}
\end{figure}

Fig.~\ref{goodwSCLC} shows more examples of the category ``good set with SCLC", and Fig.~\ref{goodwoSCLC} for ``good set without SCLC". Fig.~\ref{dist}, on the other hand, provides the distributions of the reset voltages and the SCLC onset voltages in the good-set group, and the probability of observing the SCLC feature in an $I$--$V$ curve when the $I$--$V$ curve exhibits the respective reset voltage. It can be seen that both the resets and the SCLC onsets occur most often at voltages between $-$0.6 and $-$0.7 V, and in this reset-voltage interval the probability of \textit{not} observing SCLC also reaches a local maximum. This may imply that some of the resets are triggered by SCLC, especially for the filaments with less stable atomic configurations. As the SCLC-dominating regime kicks in between $-$$0.6$ and $-$$0.7$ V in most cases, it may trigger a reset process, which leads to a tendency for the filament to break at a voltage between $-$$0.6$ and $-$$0.7$ V, being unable to exhibit the SCLC feature in the $I$--$V$ curve. This results in a local maximum of the ``Prob. w/o SCLC" line in Fig.~\ref{dist}. It is speculated that the trap-filling events or the abrupt involvement of SCLC in the conduction can affect filaments with less stable atomic configurations. More research is needed to fully understand the mechanism.

\subsection{Fair Set}

\begin{figure}[!htb]
\centering
\includegraphics[width=1.15\textwidth,center]{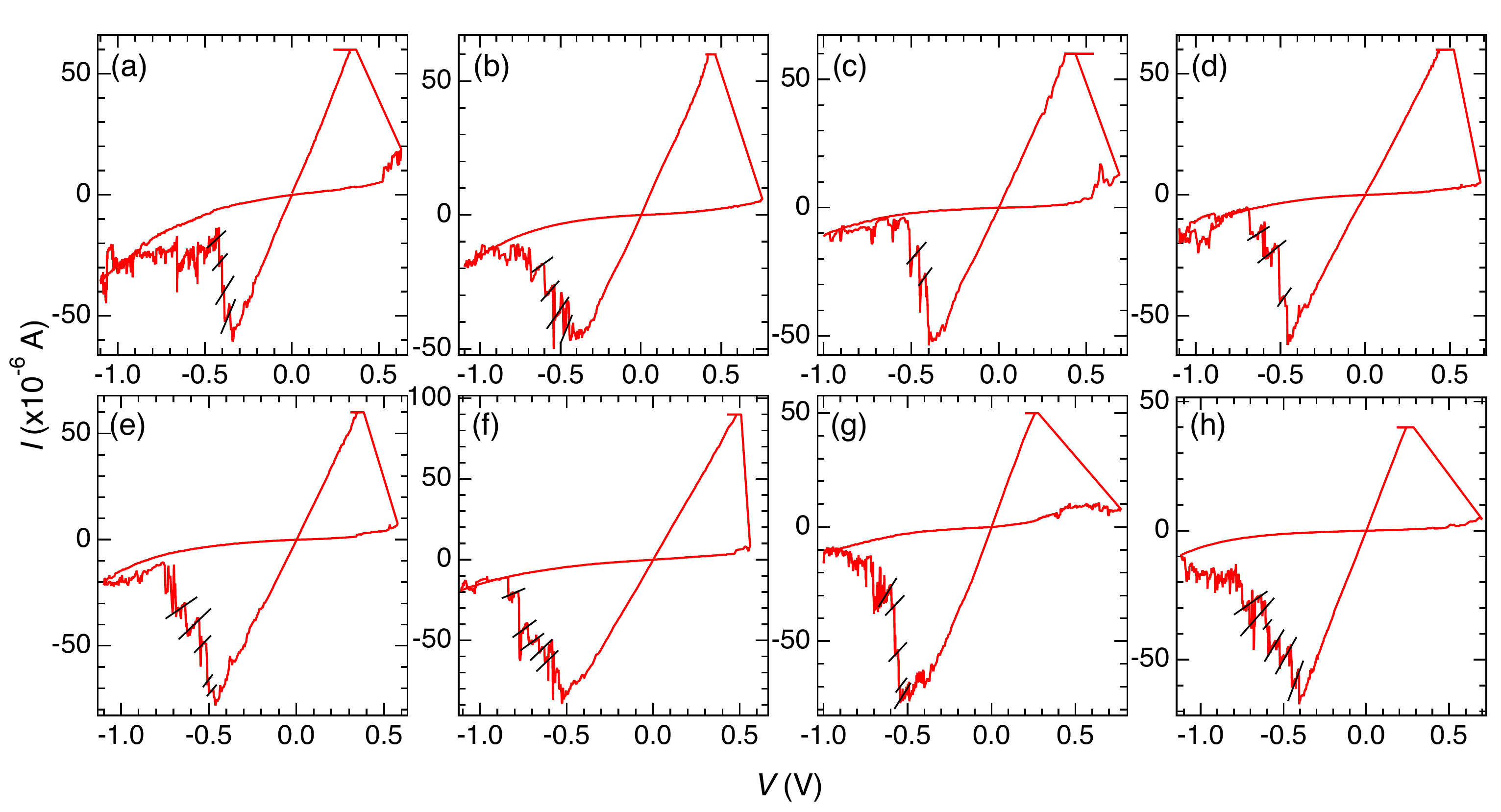}
\caption{Representative $I$--$V$ curves of fair sets. (a)–(e):  $I_\mathrm{c}$ = 60 $\upmu$A. (f)  $I_\mathrm{c}$ = 90 $\upmu$A (same data as Fig.~3c). (g) $I_\mathrm{c}$ = 50 $\upmu$A (same data as Fig.~3b). (g) $I_\mathrm{c}$ = 40 $\upmu$A.}
\label{fair}
\end{figure}

Fig.~\ref{fair} shows more examples of the fair-set category with various $I_\mathrm{c}$. It can be seen from Table 1 that the set $I_\mathrm{c}$ that can yield conductance quantization in the reset procedure falls between 40 and 90 $\upmu$A. The current at which a reset process starts for all the measurements with conductance quantization also stays between 40 and 90 $\upmu$A, making the $I$--$V$ curves exhibit a more symmetric image on the set and reset sides than the good-set curves.

\subsection{Poor Set} 

\begin{figure}[!htb]
\centering
\includegraphics[width=0.7\textwidth,center]{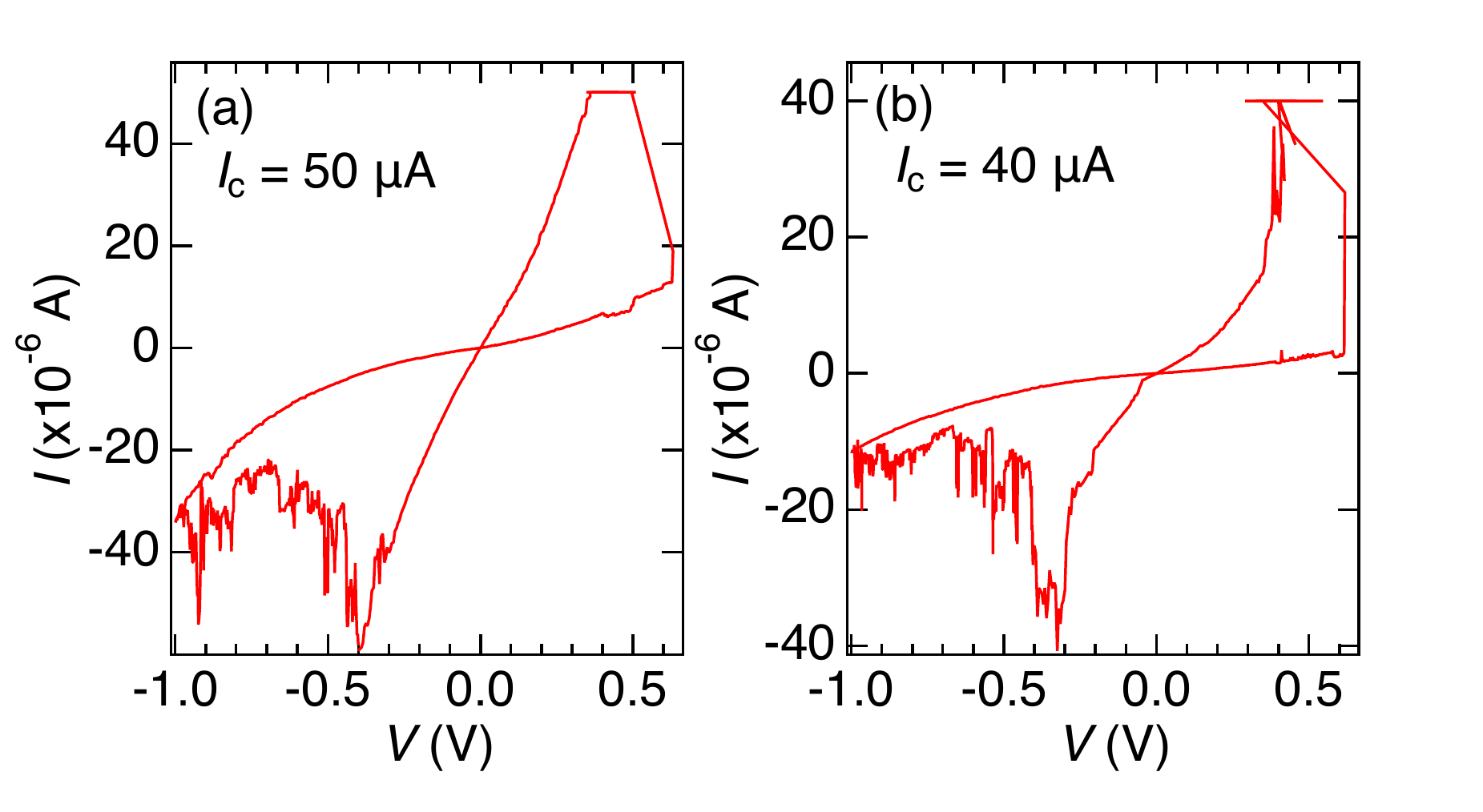}
\caption{Representative $I$--$V$ curves of poor sets.}
\label{poor}
\end{figure}

Fig.~\ref{poor} shows more examples of the poor-set category.

\section{Measurements of Devices with Different Cell Areas}\label{sec:cell}

We have fabricated dozens of devices of the same structure with the same or different cell areas, and all of them show similar behaviors in the measurements, in particular the tendency to exhibit fair sets at lower $I_\mathrm{c}$, and the high yield rate of conductance quantization during the reset procedure when a fair set is detected. Tables \ref{device60} and \ref{device96} are two representative examples of the statistics of the electrical characteristics of devices with a cell area of 0.60 $\upmu$m$^2$ and of 0.96 $\upmu$m$^2$, respectively. The measurement for each device (including the device presented in the paper with the statistics summarized in Table 1) starts from the lowest $I_\mathrm{c}$ ($=40$ or 30 $\upmu$A) to the highest ($=165$ $\upmu$A) with 15 cycles for each $I_\mathrm{c}$, followed by another 15 cycles repeated for each $I_\mathrm{c}$ starting from the highest $I_\mathrm{c}$ this time to the lowest. The statistics from the second 15 cycles have always looked very similar to the first 15, which is consistent with the hypothesis that the average diameter of the filament grown depends mostly on $I_\mathrm{c}$. Therefore, we simply display the total statistics from the 30 cycles for each $I_\mathrm{c}$ in the tables.

Although the optimal $I_\mathrm{c}$ for conductance quantization may show a little device-to-device variability, the $I$--$V$ characteristics do not seem to have apparent dependence on the cell area.

\begin{table}[!htb]
\small
\centering
\setlength{\tabcolsep}{0.29em} % for the horizontal padding
{\renewcommand{\arraystretch}{1.1}% for the vertical padding
\begin{tabular}{lrrrrrrrrrrrr}
\hline
$I_\mathrm{c}$ ($\upmu$A) & 165 & 150 & 135 & 120 & 105 & 90 & 80 & 70 & 60 & 50 & 40 & 30 \\ 
\hline
a) Good set                 & 30  & 30  & 30  & 30 & 30 & 24 & 24  & 15  & 7  & 5  & 4 & 1  \\ 
\rowcolor{Gray}
{\,\,\,\,\,\,\,w/ SCLC}      & {22}  & {18}  & {17}  & {20} & {20} & {10} & {11}  & {5}  & {2}  & {2}  & {0}  & {0}\\ 
b) Fair set                 & 0 & 0 & 0 & 0 & 0 & 6  & 6   & 15   & 22   & 22   & 20 & 14   \\ 
\rowcolor{Gray}
{\,\,\,\,\,\,\,w/ \textbf{Quant.}}      & {0}  & {0}  & {0} & {0} & {0} & {5} & {5}  & {15}  & {17}  & {19}  & {18} & {8} \\ 
c) Poor set                 & 0  & 0  & 0  & 0  & 0  & 0  & 0   & 0   & 1   & 3   & 6  & 14 \\ 
d) Set failure               & 0  & 0  & 0  & 0  & 0  & 0  & 0   & 0   & 0   & 0   & 0  & 1 \\
\hline
\end{tabular}
}
\caption{Statistics from a device with a cell area of 0.60 $\upmu$m$^2$.}
\label{device60}
\end{table}

\begin{table}[!htb]
\small
%\centering
\setlength{\tabcolsep}{0.29em} % for the horizontal padding
{\renewcommand{\arraystretch}{1.1}% for the vertical padding
\hskip-0.6cm\begin{tabular}{lrrrrrrrrrrrr} 
\hline
$I_\mathrm{c}$ ($\upmu$A) & 165 & 150 & 135 & 120 & 105 & 90 & 80 & 70 & 60 & 50 & 40 \\ 
\hline
a) Good set                 & 30  & 30  & 30  & 30 & 29 & 22 & 20  & 11  & 6  & 4  & 0 \\ 
\rowcolor{Gray}
{\,\,\,\,\,\,\,w/ SCLC}      & {20}  & {20}  & {21}  & {17} & {17} & {15} & {8}  & {3}  & {3}  & {1}  & {0} \\ 
b) Fair set                 & 0 & 0 & 0 & 0 & 1 & 8  & 9   & 18   & 20   & 17   & 16 \\ 
\rowcolor{Gray}
{\,\,\,\,\,\,\,w/ \textbf{Quant.}}      & {0}  & {0}  & {0} & {0} & {0} & {6} & {7}  & {15}  & {19}  & {14}  & {12} \\ 
c) Poor set                 & 0  & 0  & 0  & 0  & 0  & 0  & 1   & 1   & 4   & 8   & 12 \\ 
d) Set failure               & 0  & 0  & 0  & 0  & 0  & 0  & 0   & 0   & 0   & 1   & 4 \\
\hline
\end{tabular}
}
\caption{Statistics from a device with a cell area of 0.96 $\upmu$m$^2$.}
\label{device96}
\end{table}

\section{Temperature-Dependent Measurements}\label{sec:temp}

\begin{figure}[!htb]
\centering
\includegraphics[width=0.57\textwidth,center]{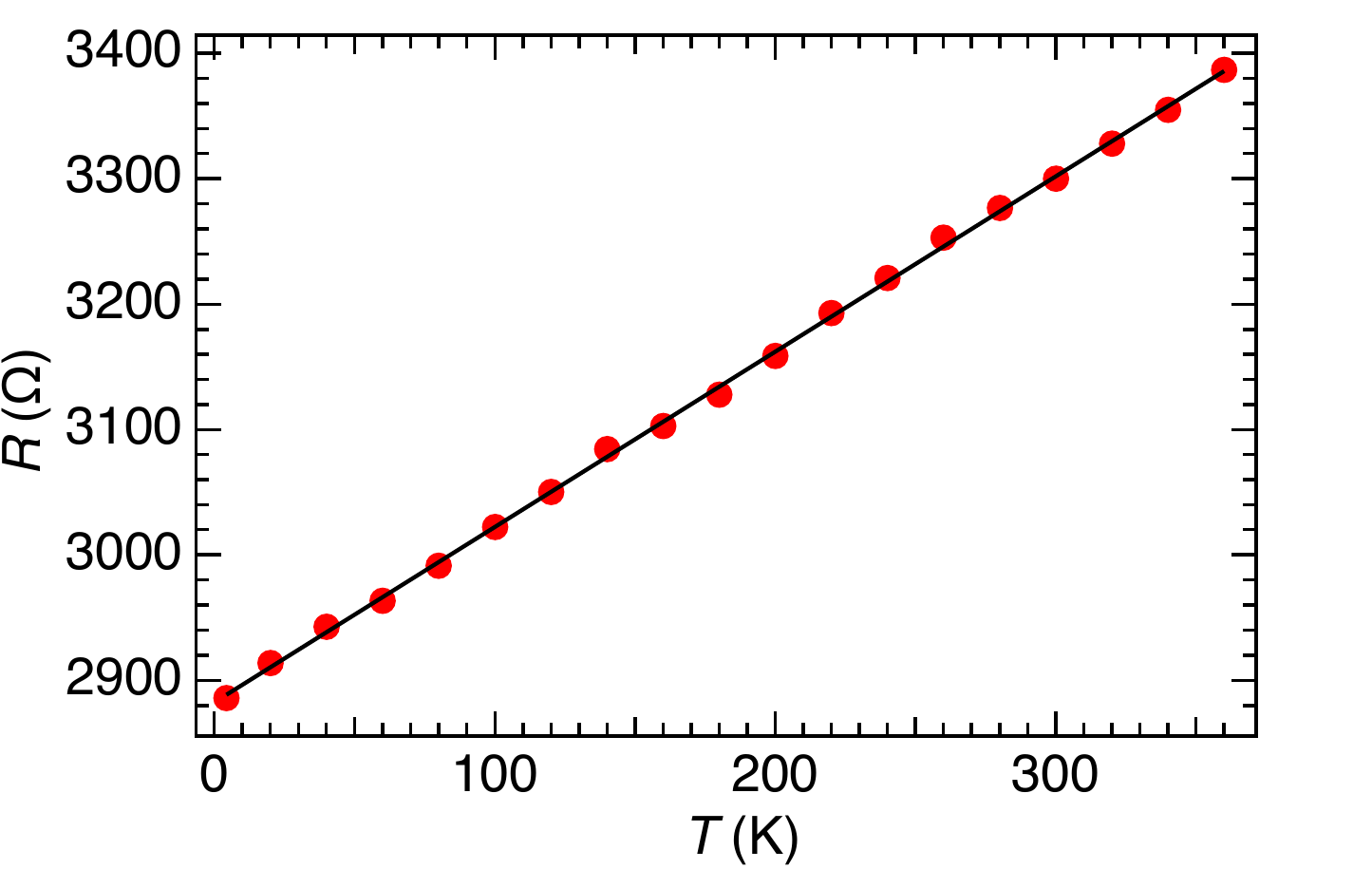}
\caption{LRS resistance as a function of temperature. The black line shows the linear fit to the curve.}
\label{RvsT}
\end{figure}

The resistance of the ohmic LRS of a filament always exhibits a decreasing trend when the ambient temperature decreases. This is a typical electronic transport behavior for a metal, in which phonon scattering is dominant. The temperature dependence of metallic resistance can be approximated as $R(T)=R_0[1+\alpha(T-T_0)]$, where $R_0$ is the resistance at temperature $T_0$, and $\alpha$ is the temperature coefficient of resistance. An example of the LRS resistance as a function of the ambient temperature is shown in Fig.~\ref{RvsT}, which is measured from a filament that has been set at a temperature of 4.2 K at $I_\mathrm{c}=135$ $\upmu$A with a good set. The resistance is read using a very small current of 10 nA with a lock-in technique to avoid excessive Joule heating at low temperature. By choosing $T_0$ as 300 K, $\alpha$ is further calculated to be $4.2 \times 10^{-4}$ K$^{-1}$. The temperature dependence of the filament resistance is small, as also found in other studies on conductive filaments in HfO$_2$-based devices \cite{Walczyk2011}.

\begin{figure}[!htb]
\centering
\includegraphics[width=0.8\textwidth,center]{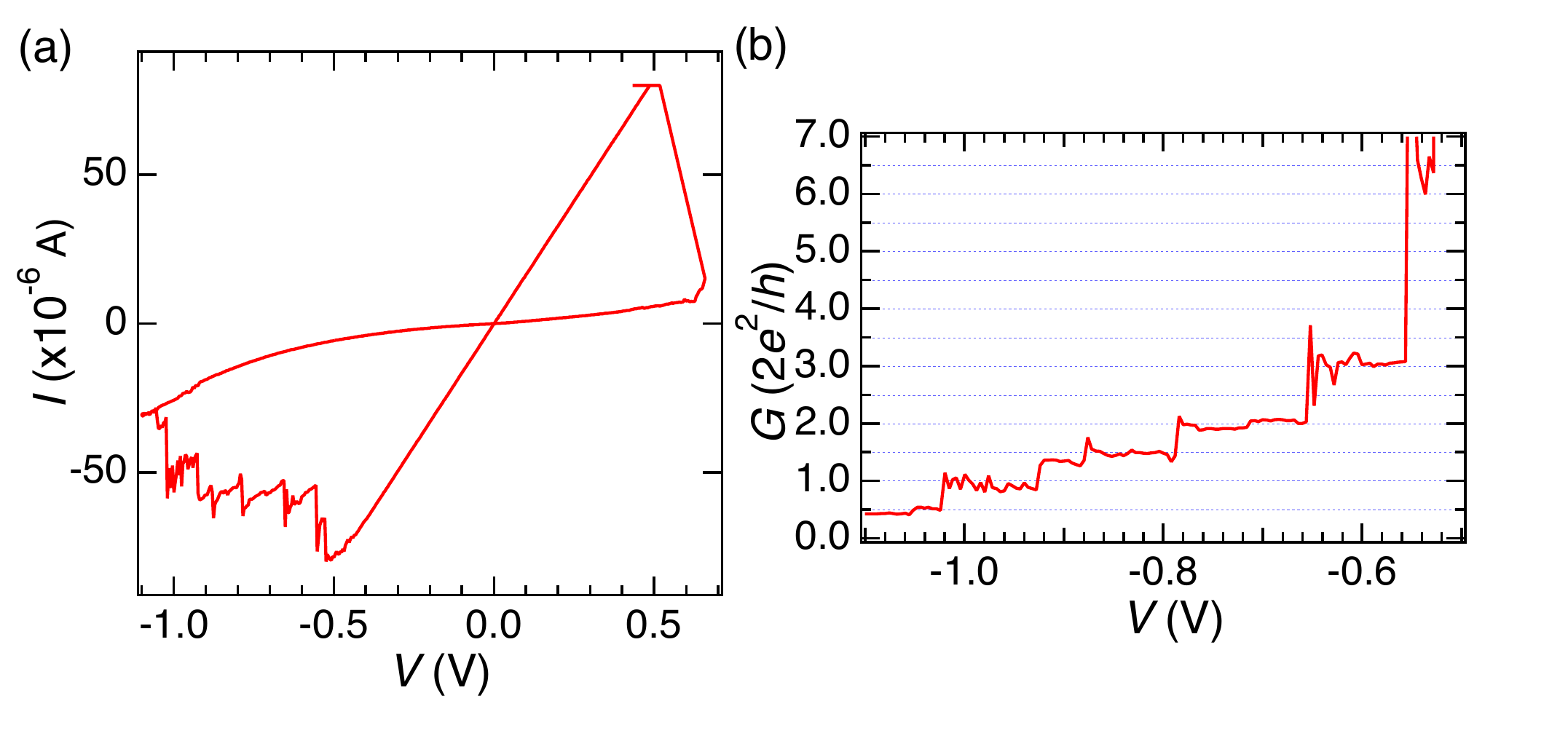}
\caption{(a) An $I$--$V$ example at 4.2 K, with its quantized conductance during the reset procedure expressed in terms of $G_0$ ($= 2e^2/h$) in Fig.~(b). The set $I_\mathrm{c}$ is 80 $\upmu$A. This plot features a wiggling set $V_\mathrm{m}$, and an ohmic LRS with a resistance of 6.1 k$\Omega$. It falls into the ``fair set" category.}
\label{IV_lowtemp}
\end{figure}

The temperature dependence of the characteristics of the set-and-reset cycles is also studied. An example of the $I$--$V$ curve at 4.2 K with a fair set and quantized conductance during the reset procedure is displayed in Fig.~\ref{IV_lowtemp}. The electrical characteristics at different temperatures from 4.2 K to 360 K look similar to those at room temperature, but statistics from these measurements show that the average reset voltage in each set category increases slightly in magnitude as temperature decreases, and becomes 0.4--0.7 V larger in magnitude at 4.2 K than that at room temperature. The statistics of the set categories at 4.2 K, on the other hand, are listed in Table \ref{lowtemp}. It can be seen that the probability of observing fair sets with quantized conductance is substantially increased especially for larger $I_\mathrm{c}$'s, and the optimal $I_\mathrm{c}$ to observe conductance quantization is increased from 60 $\upmu$A at room temperature to 80 $\upmu$A at 4.2 K.

\begin{table}[h]
\small
\centering
\setlength{\tabcolsep}{0.29em} % for the horizontal padding
{\renewcommand{\arraystretch}{1.1}% for the vertical padding
\begin{tabular}{lrrrrrrrrrrr}
\hline
$I_\mathrm{c}$ ($\upmu$A) & 165 & 150 & 135 & 120 & 105 & 90 & 80 & 70 & 60 & 50 & 40 \\ 
\hline
a) Good set                 & 15  & 13  & 13  & 14 & 12 & 10 & 5  & 3  & 0  & 0  & 0  \\ 
\rowcolor{Gray}
{\,\,\,\,\,\,\,w/ SCLC}      & {9}  & {8}  & {7}  & {7} & {5} & {5} & {2}  & {0}  & {0}  & {0}  & {0}  \\ 
b) Fair set                 & 15 & 17 & 17 & 16 & 18 & 19  & 24   & 20   & 17   & 12   & 5   \\ 
\rowcolor{Gray}
{\,\,\,\,\,\,\,w/ \textbf{Quant.}}      & {13}  & {14}  & {15} & {16} & {16} & {19} & {23}  & {18}  & {16}  & {10}  & {3}  \\ 
c) Poor set                 & 0  & 0  & 0  & 0  & 0  & 1  & 1   & 6   & 10   & 12   & 17   \\ 
d) Set failure               & 0  & 0  & 0  & 0  & 0  & 0  & 0   & 1   & 3   & 6   & 8   \\
\hline
\end{tabular}
}
\caption{Statistics about the set-and-reset cycles of the device (cell area = 0.36 $\upmu$m$^2$, same device as that presented in the paper) at 4.2 K.} 
\label{lowtemp}
\end{table}

The temperature dependence of the device characteristics can be interpreted with the following mechanisms. First, the mobility of oxygen vacancies is dependent on temperature \cite{Singh2018}. Lower mobilities at lower temperatures may explain the shift of the optimal $I_\mathrm{c}$ from 60 $\upmu$A to 80 $\upmu$A, as the filament growth and dissolution dynamics is highly dependent on the ion or vacancy mobility \cite{Yang2013a}. According to our experimental results, it is also speculated that a lower mobility and diffusivity at lower temperature is favorable for development of multiple metastable states of a filament, leading to the increased probability of observing fair sets (i.e., wiggling $V_\mathrm{m}$) especially for larger $I_\mathrm{c}$'s. The second key mechanism is that Joule heating can be better dispersed at lower temperatures \cite{Zhang2020}, especially at a temperature as low as 4.2 K. This, together with the lower mobility of oxygen vacancies, may explain the higher reset voltage and the higher yield of progressive resets (instead of abrupt ruptures) with quantized conductance at 4.2 K than at room temperature. A very low ambient temperature is therefore considered optimal for the devices to exhibit conductance quantization.

\begin{figure}[t!]%[!htb]
\centering
\includegraphics[width=0.6\textwidth,center]{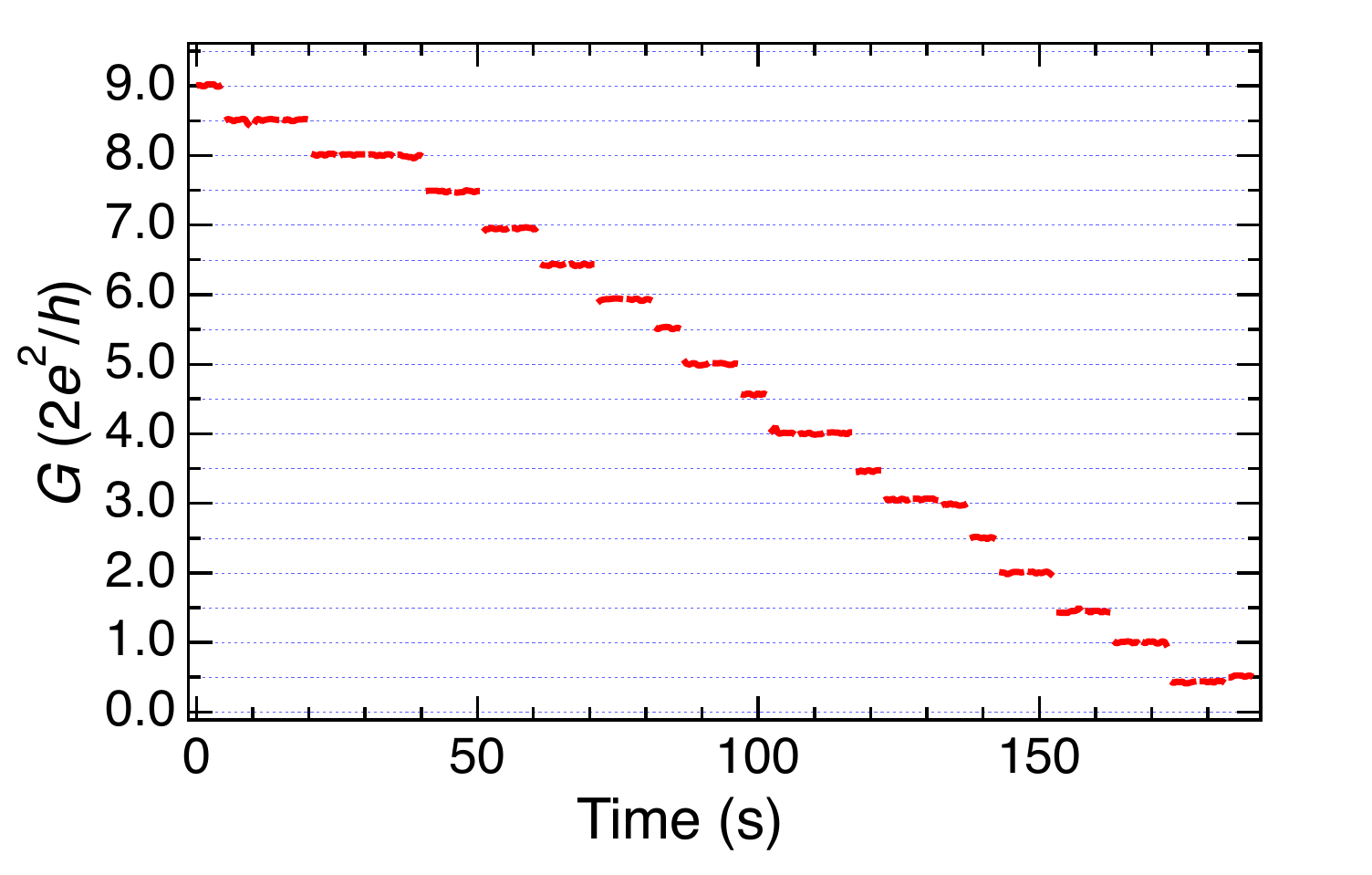}
\caption{A representative example at 4.2 K of controllable quantized conductance states at integer multiples of $0.5G_0$ using a pulse-mode reset procedure after a fair set with $I_\mathrm{c}$ = 80 $\upmu$A.}
\label{pulse_lowT}
\end{figure}

The device is also operated with the pulse-mode reset procedure at 4.2 K, with a representative result shown in Fig.~\ref{pulse_lowT}, which looks very similar to the result at room temperature (Fig.~6(a)). To limit the number of voltage stimuli within 4 for each conductance plateau, voltage pulses with fixed width of 0.1 second and fixed value of $-0.41$ V are found optimal and thereby used at 4.2 K. It has a larger magnitude than the voltage stimuli used at room temperature (i.e., $-0.35$ V). This is consistent with the higher reset voltages observed at lower temperature in dc voltage sweep mode. The signal-to-noise ratio shows very little dependence on temperature. The average standard deviation is $\sim$$0.012G_0$ for the quantized conductance plateaus at 4.2 K, which is close to that at room temperature. This confirms the picture of local heating by recurrent pulsed voltage stimuli as the main source of noise during a reset procedure \cite{Yi2016}, as the filament is being ruptured atomically under Joule heating that results in a much higher local temperature than room temperature \cite{Dirkmann2018}.

\section{The Training Sequence}\label{sec:training}

\begin{figure}[!htb]
\begin{minipage}[c]{0.6\textwidth}
\includegraphics[width=1\textwidth]{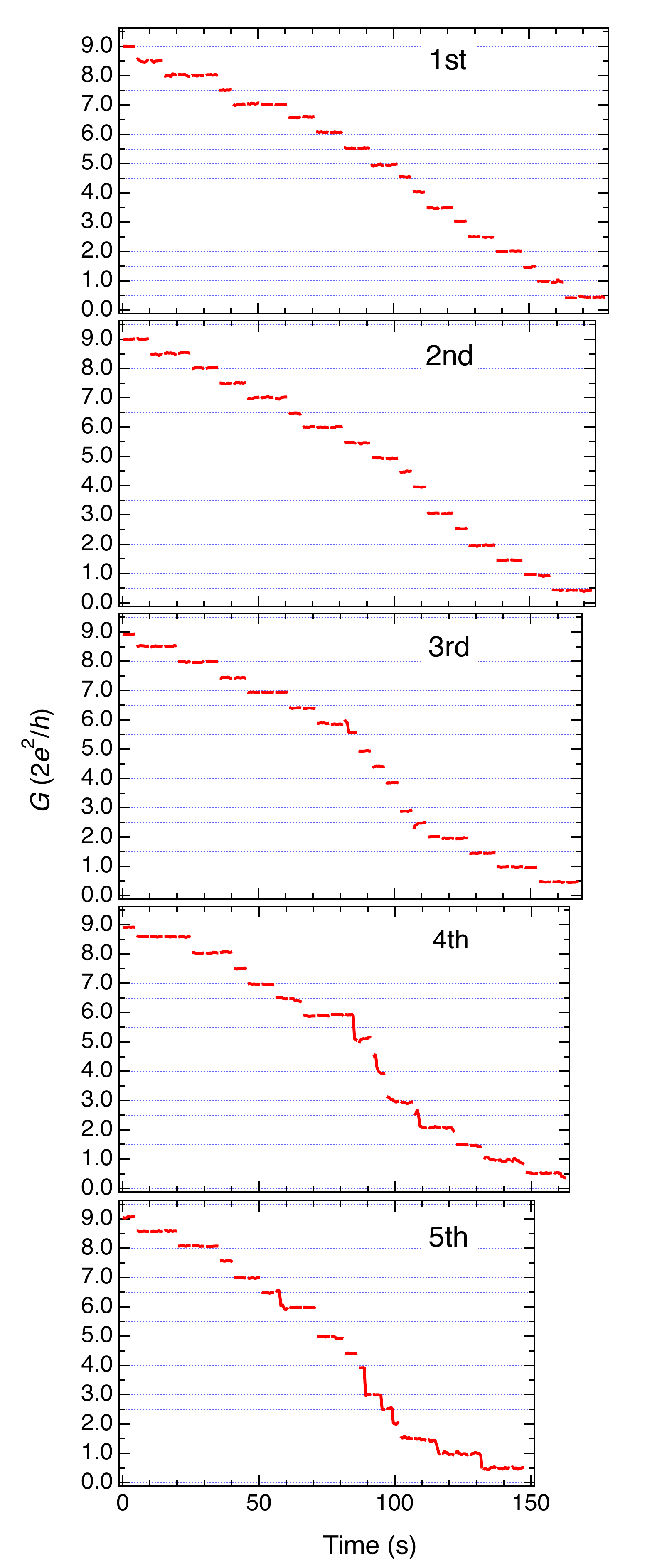}
\end{minipage}\hfill
\begin{minipage}[c]{0.4\textwidth}
\caption{A complete sequence of the training effect using the pulse-mode reset procedure.}
\label{training}
\end{minipage}
\end{figure}

The complete sequence of the training effect shown in Figs.~6a and 6b is displayed in Fig.~\ref{training}.
 
\section{Operation Using a Pulse-Mode \textit{Set} Procedure}

HfO$_2$-based devices can also undergo a \textit{set} procedure using a pulsed-voltage mode to form an ohmic conductive filament \cite{Wang2015,Yuan2017,Lin2017,Mahata2020,Tang2020,Zazpe2014}. Although being capable of yielding progressive conductance decrease in the reset procedure, memristors with high-$\kappa$ insulating layers like HfO$_2$ tend to undergo an abrupt set (under voltage stresses in either a dc sweep mode or a pulse mode) owing to the local field enhancement factor \cite{Panda2018, McPherson2003}. Therefore, it is not easy to find quantized conductance in the set procedure. Some have reported observation of conductance quantization in the set procedure, but the number of quantization plateaus is limited to only one during each set procedure, and the set process is still abrupt in nature \cite{Ulhas_thesis}. We can successfully perform a set procedure on our device either with a dc sweep mode or with a pulse mode (the voltage required for a successful set is usually $\sim$0.5 V higher in a 0.1-sec pulse mode than in a dc sweep mode), but we have not observed any conductance quantization in the set procedure. For our devices that are in a condition with potential to yield quantized conductance during the reset process, the set procedure would exhibit a wiggling $V_\mathrm{m}$ (which implies multiple metastable states) instead of conductance plateaus, as stated in the paper.

We have also tried to find quantized conductance on the reset side after a pulse-mode set procedure, but the success rate is very low ($\lesssim$15\% using optimal $I_\mathrm{c}$). Most showed an abrupt reset, and a few showed a gradual reset but with noise-like fluctuations that do not correspond to (half-)integer multiples of $G_0$ (i.e., similar to the reset behavior in the poor-set category). We suspect that a short but $\sim$0.5 V-higher voltage stress applied during a pulse-mode set procedure is not a favorable condition for developing multiple metastable states of the atomic configuration of a filament, which is known to be a requirement for yielding quantized conductance in the reset procedure for our devices.

\bibliography{atomic_9-acs-SI_c}
\newpage